# Maintaining the validity of inference from linear mixed models in stepped-wedge cluster randomized trials under misspecified random-effects structures

Yongdong Ouyang[1,2], Monica Taljaard[1,2], Andrew B Forbes[3], Fan Li[4,5]

[1] Clinical Epidemiology Program, Ottawa Hospital Research Institute, 1053 Carling Ave, Ottawa, ON, Canada

[2] School of Epidemiology and Public Health, University of Ottawa, 600 Peter Morand Crescent, Ottawa, ON, Canada

[3] School of Public Health and Preventive Medicine, Monash University, Melbourne, Australia

[4] Department of Biostatistics, Yale School of Public Health, New Haven, CT, USA

[5] Center for Methods in Implementation and Prevention Science, Yale School of Public Health, New Haven, CT, USA

**Corresponding Author:**

Dr. Yongdong Ouyang

1053 Carling Avenue, Ottawa, ON, Canada

Phone: 613-798-5555 ext. 18618

Email: yongdongouyang12@gmail.com

# Abstract

Linear mixed models are commonly used in analyzing stepped-wedge cluster randomized trials (SW-CRTs). A key consideration for analyzing a SW-CRT is accounting for the potentially complex correlation structure, which can be achieved by specifying random-effects. The simplest random effects structure is random intercept but more complex structures such as random cluster-by-period, discrete-time decay and more recently, the random intervention structure, have been proposed. Specifying appropriate random effects in practice can be challenging: assuming more complex correlation structures may be reasonable but they are vulnerable to computational challenges. To circumvent these challenges, robust variance estimators (RVE) may be applied to linear mixed models to provide consistent estimators of standard errors of fixed effect parameters in the presence of random-effects misspecification. However, there has been no empirical investigation of RVE for SW-CRT. In this paper, we review six RVEs (both standard and small-sample bias-corrected RVEs) that are available for linear mixed models in R, and then describe a comprehensive simulation study to examine the performance of these RVEs for SW-CRTs with a continuous outcome under different data generators. For each data generator, we investigate whether the use of a RVE with either the random intercept model or the random cluster-by-period model is sufficient to provide valid statistical inference for fixed effect parameters, when these working models are subject to random-effect misspecification. Our results indicate that the random intercept and random cluster-by-period models with RVEs performed adequately. The CR3 RVE (approximate jackknife) estimator, coupled with the number of clusters minus two degrees of freedom correction, consistently gave the best coverage results, but could be slightly conservative when the number of clusters was below 16. We summarize the implications of our results for linear mixed model analysis of SW-CRTs and offer some practical recommendations on the choice of analytic model.

# 1. Introduction

The cluster randomized trial (CRT) is a commonly used design for evaluating the effectiveness of cluster-level interventions or when there are substantial risks of contamination of the intervention within clusters.[1] In CRTs, individuals who are members of the same group (such as a neighbourhood, hospital, or school) are randomized as an entire unit to either the control or the intervention arms. A major implication of cluster randomization is that individuals from the same cluster tend to be more similar than individuals from different clusters. This similarity is measured by the intracluster correlation coefficient (ICC).[2] In recent years, a novel type of longitudinal CRT design, referred to as the stepped-wedge cluster randomized trial (SW-CRT) has become increasingly popular.[3,4] In the basic version of this design (Figure 1), all clusters commence the trial in the control condition and conclude in the intervention condition, with clusters randomized to one of several treatment sequences defined by the time of transition from the control condition to the intervention condition. Depending on the sampling structure, a SW-CRT can be either cross-sectional (a different set of individuals is included in each observation period), closed cohort (each individual contributes to all observation periods) or open cohort.[3] Each observation period (e.g., each cell in Figure 1) is also referred to as a cluster-period.

A commonly used analytical model for SW-CRTs is the linear mixed model,[5] which accounts for the ICC through the specification of random effects. The specification of random effects induces particular within-cluster correlation structures for the marginal distribution of the outcome vector for each cluster. Three popular forms of correlation structures induced by specification of random effects are the exchangeable (EXCH),[4] nested exchangeable (NE),[6,7] and discrete time decay (DTD) structures.[8] Sample size and power calculation software are available for SW-CRTs.[9–12] The choice of different random-effects structures and hence the correlation structures could impact both sample size calculations and statistical inference.[5,13,14] The exchangeable model arises from a single random intercept for clusters and implies a common ICC both within and across periods.[2] The nested exchangeable model arises from a model with a random intercept and an additional random cluster-by-period interaction effect. This correlation structure allows researchers to distinguish a within-period ICC (correlation between measurements from two individuals in the same cluster-period) and a constant between-period ICC (correlation between measurements from two individuals in the same cluster but different periods).[6] The ratio of the within- and between-period ICCs is defined as the cluster autocorrelation coefficient (CAC) and is generally smaller than one (i.e., when the variance component for the random cluster-by-period interaction is greater than zero).[6] Finally, the DTD model assumes that the between-period ICC decays exponentially by period and arises from a first-order autoregressive (AR(1)) random-effects specification.[8] In this model, the CAC measures the decay rate per period.[8] In addition to these three types of correlation structures, it has also been recommended to allow for cluster-treatment effect heterogeneity, i.e., to allow the treatment effect to vary across clusters by including a random cluster-by-intervention interaction effect.[15,16]

The existing literature has demonstrated the potential impact of random-effects misspecification in linear mixed models and generalized linear mixed models. Frequently, the random effects are assumed to follow normal (Gaussian) distributions, which can be incorrectly specified when their true distributional forms deviate from normality.[17] Litière et al.[18] showed that under an individually randomized design, error distribution misspecifications can either over- or under-estimate power and could substantially inflate the Type I error rate. In SW-CRTs, rather than error distribution misspecification, random-effects misspecification can also occur when certain random effects are omitted (e.g., assuming an exchangeable correlation structure when a decay parameter is needed).[19] In this article, we focus on this latter type of misspecification and empirically explore methods that can maintain the validity of inference under such misspecification. In practice, investigators are often faced with an array of modeling choices without clear guidance as to which random effects or correlation structures they should choose. When the true random effects component is unknown, the final choice may depend heavily on subject-matter expertise and/or subjective judgment. Hui et al.[20] empirically demonstrated the sensitivity of linear mixed model inference to the random effects choices with clustered data. Voldal et al.[21] examined implications of misspecified correlation structures in SW-CRTs (with linear mixed models) and showed that misspecification affects both Type I error rates and efficiency of the treatment effect estimator. Although some guidance has recently been provided on selection of a correlation structure using information criteria for continuous outcomes, adequate selection performance requires a large number of clusters, and the validity of post-selection inference in SW-CRTs has not been sufficiently investigated.[22] In fact, since the true correlation structure for any application is always unknown, it is also impossible to assess the impact of correlation structure misspecification for any given application.[23,24] In addition, even if researchers believe a more complex model is appropriate, it may not be possible to fit a more complex model, as SW-CRTs often have a small number of clusters[25,26] and computational challenges may arise.

A potential solution to misspecification has been provided under the generalized estimating equations (GEE) framework,[27] where researchers may utilize the robust variance estimator (RVE), sometimes referred to as the sandwich variance estimator, to consistently estimate the sampling variance of the treatment effect estimator in an assumed marginal model for the mean of the outcome.[28–30] Although RVEs have been implemented in mixed-effects models and previous studies have hinted that RVEs might be used to maintain validity of inferences under misspecified random-effects structures,[21,31] the performance of RVEs in mixed-effects models has not been empirically examined in the context of SW-CRTs, and there has been no recommended practice for their use to date. Given the popularity of the mixed-effects model in SW-CRTs,[5,32] we investigated the performance of RVEs under different types of misspecified random-effects structures. Furthermore, as SW-CRTs often have a limited number of clusters, and several modified versions of RVEs with small-sample corrections have been proposed, we also explored the performance of small-sample corrections that are currently available in standard software for mixed-effects models — in particular, the R software package. Therefore, our objectives were to conduct an extensive simulation study to address the following emerging questions:

- First, *does the use of RVEs in linear mixed models lead to valid inferences for the estimation of the intervention effect when the random-effects structure is misspecified, even in the most severe form of misspecification, i.e., with a simple exchangeable correlation*?
- Second, *if small sample corrections are required to maintain statistical properties with a small number of clusters, which available small-sample corrections in linear mixed model perform the best when using RVEs under random-effects misspecification?*

The remainder of the manuscript is organized as follows. In Section 2, we review the RVEs that are currently available in the R package "clubSandwich" for mixed-effects models. In Section 3, we present the results of a simulation study to investigate the performance of these RVEs for SW-CRT across a range of scenarios. In Section 4, we demonstrate the use of RVEs in a real data example. Section 5 concludes with a discussion.

## 2. Robust variance estimators and small-sample corrections in linear mixed models

### 2.1. Model and Notations

We review some general notation and linear mixed model formulations in the context of the multilevel data structure. Commonly used linear mixed models in SW-CRTs are special cases of this general model formulation.[5] Let $Y_i$ be a vector of outcomes for individuals from the $i$-th cluster ($i = 1, \ldots, I$) and let $X_i$ be a design matrix of corresponding covariates. The vector $Z_i$ is the design matrix of random effects for each cluster. A generic linear mixed model can be written as:

$$Y_i = X_i \beta + Z_i u_i + \epsilon_i$$

(1)

where $\beta$ is the fixed-effects regression parameters, $u_i$ follows the multivariate normal distribution $N_q(0, R)$, where $q$ is the dimension of the random-effects parameters, and $R$ is a $q \times q$ random-effects variance matrix. In a SW-CRT, $\beta$ typically includes the time effects

(secular trend) and intervention effect parameters. Specific examples of the random-effects structure parameterized by $R$ will be provided in Section 3. In addition, $\epsilon_i = (\epsilon_{i1}, \ldots, \epsilon_{in_i})'$ is a vector of random errors for all individuals in the $i$-th cluster, and is assumed to follow a normal distribution $N_{n_i}(0, \sigma_\epsilon^2 I_{n_i})$, where $n_i$ is the cluster size (number of observations made in the $i$-th cluster across all periods), $\sigma_\epsilon^2 = Var(\epsilon_{ij})$ and $I_{n_i}$ is the $n_i \times n_i$ identity matrix. The total variance for each cluster then can be denoted as

$$V_i = Var(Y_i) = Z_i R Z'_i + \sigma_\epsilon^2 I_{n_i}$$

(2)

If the variance components are known, the estimated fixed effects can be obtained from the generalized least squares formula:

$$\hat{\beta} = (X'V^{-1}X)^{-1} X'V^{-1}Y$$

(3)

Where $V$ is a block-diagonal variance matrix with $V_i$ on the diagonal and zero on the off-diagonal, and $X$ is the joint design matrix across all clusters (by stacking $X_i$ over all clusters), and $Y = (Y'_1, \ldots, Y'_I)'$ is the vector of all outcomes observed in the trial. Based on the work by Liang and Zeger,[28] the cluster RVE for a linear mixed model can be written as follows:[33–35]

$$Var(\hat{\beta}) = (X'V^{-1}X)^{-1} \left( \sum_{i=1}^{I} X'_i V_i^{-1} \Sigma_i V_i^{-1} X_i \right) (X'V^{-1}X)^{-1}$$

(4)

where the true covariance matrix $\Sigma_i = Var(Y_i - X_i\beta) = Var(r_i)$ is unknown and must be estimated from the data. In particular, with a small number of clusters, there is a tendency for

$X_i\hat{\boldsymbol{\beta}}$ to be too close to the observation $Y_i$, making the estimated residuals $\hat{r}_i$ too close to zero, hence biasing variance estimates downward.[36] To address this small sample bias, $\Sigma_i$ in (4) needs to be inflated to reflect the correct uncertainty of the regression parameter estimates. There are several variations of (4) that have been developed to address small sample bias. In our study, we particularly focus on corrections that are currently available in the "clubSandwich" package in R, which is targeting the cluster RVE under mixed-effects models.[37] The available RVEs in this package can be written in the form:

$$\widehat{Var}(\hat{\boldsymbol{\beta}}) = (X'V^{-1}X)^{-1} \sum_{i=1}^{I} X'_i V^{-1} A_i \hat{r}_i \hat{r}'_i A'_i V^{-1} X_i (X'V^{-1}X)^{-1}$$

(5)

where $r_i$ is the residual vector in cluster $i$, and $A_i$ is an adjustment matrix. A standard RVE (denoted CR0) is to let $A_i = I$. For multilevel regression models in general, the standard RVE may lead to inflated Type I error if the number of independent units is fewer than 50.[38] Similar findings were discussed in the literature for GEE analysis of SW-CRTs,[27,39–41] but little evidence has been provided for mixed-model analysis of SW-CRTs. A recent review of 160 published SW-CRTs found that 70% of studies used linear or generalized linear mixed models as the primary analytical approach even though the median number of clusters was only 11.[26] Therefore, small sample corrections to the standard RVEs are often required when implementing linear mixed models in practice and we will review available small sample corrections in the next section.

**2.2. Simple small sample corrections**

A simple small sample correction is to adjust the CR0 variance estimator by different degree of freedom (DoF) corrections. In "clubSandwich", there are three types of DoF corrections: (1) correct the standard RVE by $I/(I-1)$ (denoted as CR1); (2) correct the standard RVE by $I/(I-P)$, where $P$ is the number of fixed-effects covariate parameters estimated in the model (denoted as CR1P); or (3) correct the standard RVE by $(I(N-1))/[(I-1)(N-P)]$, where $N$ is the total number of observations (denoted as CR1S).

### 2.3. Bias-reduced linearization and approximation of Jackknife estimator

In addition to the DoF correction, a common correction is to use resampling methods, such as jackknife resampling estimators.[42–44] Mancl and DeRouen[36] proposed a bias-reduced correction by deriving an approximation for the bias of $\hat{r}_i\hat{r}_i'$ with a first-order Taylor series expansion of the residuals. Bell and McCaffrey[45] showed that this correction closely approximated the jackknife estimator under an unweighted linear model. This correction, denoted as CR3, replaces $A_i r_i r_i' A_i'$ by

$$A_i r_i r_i' A_i' = [I_i - H_i]^{-1} r_i r_i' [I_i - H_i]^{-1}$$

(6)

where $H_i = X_i(X'V^{-1}X)^{-1}X_i'V_i^{-1}$ is the cluster-leverage matrix. This approach is also known as the Mancl and DeRouen variance correction in the GEE literature and has been found to be an effective small-sample method (but sometimes conservative due to over-correction) for GEE estimators in SW-CRTs.[36] Finally, Bell and McCaffrey found that CR3 tends to over-correct the bias of CR0, and thus a bias-reduced linearization (BRL) method was proposed. This correction (denoted as CR2) is close to CR3:

$$A_i r_i r_i' A'_i = [I_i - H_i]^{-1/2} r_i r_i' [I_i - H_i]^{-1/2}$$

(7)

Except that the terms parameterizing (6) and (7) were specifically for linear mixed models, this approach is also known as the Kauermann and Carroll variance correction in the GEE literature,[46] and it has been found to maintain the nominal type I error rate for GEE analyses of small SW-CRTs under different multilevel working correlation structures (but mostly under the correct specification of the working correlation structure).[39–41,47] To investigate the potential for these variance estimators to yield valid inference under misspecified linear mixed models, we conducted a simulation study to compare their finite-sample operating characteristics. For ease of referene, a summary of the RVEs and their features is provided in Table 1.

## 3. Simulation study

In this section, we report the results of a simulation study to determine whether utilizing RVEs can help to maintain valid inference when the correlation structure has been incorrectly specified in SW-CRTs and to identify which specific RVE performs best under different scenarios, including with a limited number of clusters.

### 3.1. Trial design

We considered the standard SW-CRT design with cross-sectional measurement, assuming $I$ clusters, $S$ sequences and $J = S + 1$ periods of equal length (Figure 1). We assumed balanced designs such that $I/S$ clusters are included per sequence, and equal cluster-period sizes $K$.

### 3.2. Data generating process

We assumed a continuous outcome and generated data under two different models: (1) the Discrete Time Decay with Random Intervention (DTD-RI) model and (2) the Nested Exchangeable with Random Intervention (NE+RI) model. Both data generators share the same mean model (i.e., fixed effects) and only differ in their random effect structures. While we explicitly write out each data generator below, we refer to Li et al.[5] for a unified model representation that includes these two data generators as special cases with different specifications of the random-effects structure.

(1) Discrete Time Decay with Random Intervention (DTD-RI) model

The DTD-RI model can be written as

$$Y_{ijk} = u + \beta_j + (\theta + v_i)X_{ij} + \gamma_{ij} + \varepsilon_{ijk}$$

where $Y_{ijk}$ denotes the outcome of the $k$-th individual from the $i$-th cluster and $j$-th period; $\beta_j$ denotes the fixed time effect ($\beta_j = j$, where $j = 1, 2, \ldots, J$); $u$ is the overall mean under the control condition; $\theta$ denotes the time-invariant treatment effect, and $X_{ij}$ a binary treatment indicator for the $i$-th cluster in the $j$-th period (taking values 1 or 0 depending on the treatment condition). The individual error term $\varepsilon_{ijk}$ is assumed to follow $N(0, \tau_\varepsilon^2)$. The random treatment effect $v_i$ is assumed to follow $N(0, \tau_v^2)$ and measures the departure of the treatment effect in each cluster relative to its overall mean $\theta$. The vector of random cluster-by-period effects is assumed to follow $\boldsymbol{\gamma}_i = (\gamma_{i1}, \ldots, \gamma_{iJ})' \sim N(0, \tau_\gamma^2 \widetilde{\mathbf{Z}})$. We particularly focus on $\widetilde{\mathbf{Z}}$ as an autoregressive (AR(1)) structure,[8] that is

$$\widetilde{\mathbf{Z}} = \begin{bmatrix} 1 & r_{12} & r_{13} & \cdots & r_{1J} \\ r_{12} & 1 & r_{23} & \cdots & r_{2J} \\ r_{13} & r_{23} & 1 & \cdots & r_{3J} \\ \vdots & \vdots & \vdots & \ddots & \vdots \\ r_{J1} & r_{J2} & r_{J3} & \cdots & 1 \end{bmatrix} = \begin{bmatrix} 1 & r & r^2 & \cdots & r^{J-1} \\ r & 1 & r & \cdots & r^{J-2} \\ r^2 & r & 1 & \cdots & r^{J-3} \\ \vdots & \vdots & \vdots & \ddots & \vdots \\ r^{J-1} & r^{J-2} & r^{J-3} & \cdots & 1 \end{bmatrix}$$

where, in the absence of the random intervention effect, $r$ in the second equation represents the common amount of between-period correlation decay per period.

(2) Nested Exchangeable with Random Intervention (NE+RI) model

The NE+RI model can be written as:

$$Y_{ijk} = u + \alpha_i + \beta_j + (\theta + v_i)X_{ij} + \gamma_{ij} + \varepsilon_{ijk}$$

This model has an additional random intercept for the cluster and is denoted as $\alpha_i$, following $N(0, \tau_\alpha^2)$. The term $\gamma_{ij}$ still denotes the random cluster-period effect but now is assumed to follow $N(0, \tau_\gamma^2)$. The two random effects in this model are assumed to be independent, including a nested exchangeable correlation structure for the outcome observations within the same cluster.

The definitions for the ICCs under both models are summarized in Table 2; also see Ouyang et al[14] for a more detailed exposition of the definition of ICCs in simpler models without the random intervention effect. Notice that, due to the inclusion of random intervention effects, the definitions of the within- and between-period ICCs are different under control and intervention conditions. Furthermore, the interpretation of the CAC under the control condition is different for models (1) and (2). In the nested exchangeable model, the CAC is fixed regardless of the distance between periods, whereas, in the exponential decay model, the CAC is the fixed decay rate per period. The original definition of the CAC, namely the ratio of the between- to within-period ICCs, is still applicable under both intervention and control conditions. We emphasize that an extra term for the random intervention effect must be added to the ICC definition under the intervention condition, and the inclusion of a random intervention effect induces between-arm heterogeneity.

### 3.3. Simulation parameters

We chose simulation parameters (i.e., number of clusters, correlation parameters and cluster sizes) based on published literature. The assumed values are summarized in Table 3. For example, Nevins et al.[26,48] conducted a review of 160 published SW-CRT and found that the median number of clusters was 11 with first and third quartiles of 8 and 18. Furthermore, Korevaar et al.[49] presented a database of estimated correlation parameters from reanalyzing multiple SW-CRTs and found that ICC (depending on the correlation structure) tended to range between 0.01 and 0.1 and a typical CAC value was 0.8. We therefore generated data with $I = 8$, 16 and 32 clusters, $S = 4$ and 8 sequences and cluster-period sizes $K = 10$ or 100. We primarily focused on evaluating the validity of the Wald test (type I error rate), thus, we assumed the time-invariant treatment effect (scaled treatment effect relative to the error variance) is zero and fixed the standard deviation of the individual error term ($\tau_\varepsilon$) at 1. We considered two combinations of within-period ICCs under the control ($\rho_0$) and intervention conditions ($\rho_1$): (0.01, 0.05), (0.05, 0.10) and (0.05, 0.15). The corresponding standard deviations of the random intervention effect ($\tau_v$) are 0.21, 0.24 and 0.35. The CAC under the control condition was assumed to be 0.8, and the corresponding values of CAC under the intervention condition can then be calculated using Table 2.

### 3.4. Data analysis and performance measures

Each simulated dataset was analyzed with (1) the correctly specified model (either DTD-RI or NE-RI), and then with two misspecified models: (2) EXCH and (3) NE, each with and without RVE. We estimated six different RVEs (see Table 1) under each mixed-effects model (described in Section 2.4) using "clubSandwich" (version 0.5.10) in the R package. In total, for each simulated dataset, there were 15 fitted models. All models were fitted using "lmerTest" in R[50] with the exception of the DTD-RI model which was fitted using "glmmTMB" in R.[51] All models were fitted with their default non-linear optimizers. As an ad hoc analysis, the NE-RI models were also fitted using bound optimization by quadratic approximation ("bobyqa") in an attempt to improve the convergence rate.[52]

The definitions of the performance measures of interest are shown in Table 4. We first calculated the bias of the estimated time-invariant treatment effects. Then, we recorded the empirical (Monte Carlo) standard error (SE) of the estimated treatment effect as well as the average SE (from models with and without RVEs) and the percentage error in the averaged model-based SE (SE estimate from a model) relative to the empirical SE. The coverage of 95% confidence intervals (CI) for the treatment effect was assessed over repeated data generations.

For each trial configuration, we simulated 2,000 datasets. This allowed us to control the Monte Carlo Standard Error (MCSE) to within $\pm 0.5\%$ for a coverage probability of 95%.[53] Convergence is a major concern when fitting complex mixed-effects models even under correct model specification. Previous simulation studies have shown a considerable proportion of non-convergence, especially when the number of clusters is small.[27] We therefore also recorded the prevalence of non-convergence for all models. The data generation and analyses were conducted on the Compute Canada High-Performance Computing Cluster with R version 4.0.2.

### 3.5. Degree-of-freedom corrections

With a small number of clusters, studies have shown the need to correct the asymptotic normal approximation for Wald tests. For all our assessments of inferential properties (e.g., 95% CI coverage), we used a $t$-test with the DoF equal to the number of clusters ($I$) minus two. This choice of DoF was originally proposed by Ford and Westgate for GEE estimators,[54] and has subsequently been shown empirically to work well in several other studies that focused on GEE estimators of SW-CRTs.[41,47,55] In addition, the DoF $I - 2$ is also easier to implement in sample size calculations compared to other DoF corrections such as Satterthwaite, which are complex and data-dependent.[56] However, for general interest, we have also provided the results for the Satterthwaite DoF correction in the Supplementary material.[57] Please note that Satterthwaite DoF under the DTD-RI model is currently not available in R; therefore, the normal approximation was used under that specific model (this, however, was not an issue for the $I - 2$ DoF).

### 3.6. Simulation Results

### 3.6.1. Bias of the estimated treatment effects

Not surprisingly, since we used linear mixed models with correctly specified models for the conditional mean outcome (fixed-effects component), the estimated treatment effects were unbiased across all trial configurations, even if the random effects structure was misspecified.[21,58] The details are presented in Supplementary material, Table S1.

### 3.6.2. Coverage probabilities of 95% confidence intervals
#### 3.6.2.1. Data generated from discrete time decay with random intervention model (DTD-RI)

We first present simulation results with data generated from the DTD-RI model. The SEs of estimated treatment effects from the different fitted models varied substantially. As expected, misspecified random-effects structures (e.g., EXCH and NE) without RVEs produced smaller (but biased) average model-based SEs, which led to the coverage probabilities being smaller than the nominal 95% level. In Figure 2, we show the coverage probabilities of 95% CI obtained from the EXCH and NE models with and without RVE, and from the true models (based on model-based SE). Each quarter-block of Figure 2 presents the 95% CI coverage for various combinations of within-period ICCs, number of sequences and cluster period sizes, together with Monte Carlo bounds around 95% coverage. For example, the first quarter of Figure 2, (the top-right four mini-panels) present comparisons between models with and without RVEs, first under exchangeable (EXCH) and then under nested exchangeable (NE) correlation structures. We only present results for ICC combinations of (0.01, 0.05) and (0.05, 0.15). The results for ICC combination (0.05 and 0.10) are presented in Supplementary Table S2. The results show that the misspecified models with model-based SE estimators were generally not able to reach 95% coverage. Both the EXCH and NE models without RVEs had similar coverage when $K = 10$, but the NE models had much better coverage, especially when $K = 100$. With a model-based SE estimator, the performance of the misspecified models was worse when the within-period ICCs were larger (this is more obvious for EXCH) or when the number of sequences was larger (e.g., fewer clusters were allocated to each sequence). The coverage probabilities of the misspecified model with a model-based SE estimator were lower with larger cluster-period sizes thus, increasing the cluster-period sizes did not improve validity of inferences.

To investigate the performance of RVEs in finite samples, Figure 2 also presents the coverage probabilities under CR0 (standard RVE) and CR3 (the RVE of approximating the leave-one-cluster-out jackknife variance estimator) respectively, compared to coverage under the true model. We chose to focus on presentation of results for these two RVEs because CR0 is the standard RVE, and CR3 had the best performance among all RVEs we investigated. In general, the coverage probabilities tended to be closer to 95% when using RVEs than without RVEs. CR0 did not always reach nominal 95% coverage, even with 32 clusters. Therefore, small sample corrections are generally required to maintain the validity of misspecified linear mixed models in SW-CRTs. Corrections with simple degree of freedom (CR1 and CR1S) led to coverage probabilities much lower than the nominal 95%. The bias-reduced linearization (CR2) had better coverage and produced decent coverage probabilities in some cases. However, it generally had underestimated SEs, leading to likely under-coverage with 16 clusters. CR1P consistently led to coverage probabilities over 96% and overestimated SEs (Supplementary material, Table S2 and Table S3). It is important to note that CR1P (with correction $I/(I - P)$) was not defined when the number of clusters and sequences were both 8 (i.e., $I = 8$, $P = 9$). CR3 had reasonable performance, leading to coverage probabilities generally closer to the nominal level and similar to that of the true model. When the number of clusters was small (e.g., 8 clusters), the coverage probabilities (based on the model-based SE) under the true model exceeded the nominal level. While the same observation applied to CR3, the coverage probabilities tended to be closer to 95% in general. Satterthwaite DoF usually led to under-coverage when the number of clusters was either 8 or 16 (Supplementary material, Table S4). We found the coverage probabilities (under Satterthwaite DoF) could be as low as 91% when the number of clusters was 8. Even with 16 clusters, we still observed scenarios with 93% coverage. In general, when we fixed other parameters, the within-period ICCs and the number of sequences had limited impact on coverage for the RVE methods, although we saw slightly better coverage when the cluster-period sizes were larger.

### 3.6.2.2. Data generated from nested exchangeable with random intervention model

The results when the data were generated from the NE-RI model are similar to those from DTD-RI. In Figure 3, we present the coverage probabilities of 95% confidence intervals obtained from the misspecified EXCH and NE with and without RVE, and the true NE-RI models. Since the NE-RI is basically the NE model with a random intervention effect, the performance of the NE model with the model-based SE estimator would be associated with the standard error of the random intervention across clusters (the NE model under the ICC combination of (0.05, 0.15) shows worse performance due to larger $\tau_v$ value).

### 3.6.3. Percentage error of model-based SE to empirical SE

### 3.6.3.1. Discrete time decay with random intervention model

Supplementary material, Table S3 summarizes the percentage errors of model-based SE to empirical SE under all models. When the data were analyzed using misspecified models without RVE, the standard errors were often underestimated by over 60%. Relative errors decreased substantially when RVEs were used. In Figure S1, we compare the percentage error of model-based SEs from CR0 and CR3 to the true models. Across simulation scenarios, the patterns were similar to those presented for coverage probabilities: the correction by CR0 was too small, and standard errors were underestimated by over 24%. CR3 had the smallest relative errors compared to other types of RVEs, although the error could still be as high as 8% when the number of clusters was only 8. The overestimate of SEs indicates a potential loss in efficiency after applying RVEs. When the number of clusters was larger (e.g., 16 or 32), the relative error was usually within 5% and could be as low as 1%. Comparing across all RVEs, CR3 had relative errors closest to the relative errors obtained by fitting the true models (based on the model-based SE).

### 3.6.3.2. Nested exchangeable with random intervention model

We observed similar results when the data were generated from the NE-RI model. In Figure S2, we present the percentage error of model-based SE to empirical SE for both true model and misspecified models with CR0 and CR3.

### 3.6.4. Prevalence of non-convergence

In our simulations, we excluded results when models did not converge. Table 5 summarizes the percentage of non-convergence for each trial configuration by model. Generally, when all other parameters remain the same, larger cluster-period sizes $K$ tend to have larger non-convergence rates whereas larger numbers of clusters tend to have smaller non-convergence rates. The proportion of non-convergence was as high as 39% for DTD-RI when the number of clusters and cluster-period sizes are both small (in the most extreme scenario we investigated). Under the default optimizer ("nloptwarp": nonlinear optimizer), non-convergence for the NE-RI models was as high as 15%. However, with bound optimization by quadratic approximation ("bobyqa" optimizer), the non-convergence rate was close to zero, suggesting that this is a promising optimization routine for fitting more complex linear mixed models in SW-CRTs. As the "nloptwarp" option is a nonlinear version of the "bobyqa" option, the better convergence rate may be explained by the simpler optimization tolerance parameters required by "bobyqa".

### 3.6.5. Additional evaluations

As suggested by reviewers, we conducted additional analysis with selected scenarios to evaluate the statistical power of linear mixed models with RVEs. In addition, we compared the performance of linear mixed model with RVE with alternative robust analytical methods (GEE and permutation test) that researchers proposed in the literature.

#### 3.6.5.1. Statistical power

As an additional set of evaluations, we compared power when using linear mixed models with CR3 and the true model under selected scenarios. Specifically, we considered 6 scenarios for each data generator (12 scenarios in total), including a combination of various factors such as the number of clusters (8 and 32), sequences (4), and ICC values ((0.01, 0.05), (0.05, 0.10), (0.05, 0.15)). The results are presented in Supplementary Table S5. We primarily focus on CR3 because it has comparable coverage probabilities to the true model and the size of the test is generally well maintained to the nominal level. We observed that the loss of statistical power in the RVE (CR3) model is minimal when the number of clusters is 32 and is less than 10% when the number of clusters is only eight.

### 3.6.5.2. Comparison with GEE and permutation tests

Both GEE and permutation tests have been used in stepped wedge trials for robust inference and to protect potential random-effects misspecification.[54,59] Under the 12 scenarios mentioned in the above section, we conducted additional simulation (Supplementary, Table S6) to show that bias and coverage probabilities under linear mixed models (CR2 and CR3) were very similar to those under GEE (KC and MD).

Similarly, we compared the coverage probabilities under linear mixed models (CR2 and CR3) with permutation test. We implemented the permutation test for SW-CRT suggested by Ren et al.[60] The test statistic was defined as the estimated treatment effect under the fitted linear mixed effect model. The permutation test was implemented as follows. We first estimated the test statistics from the observed allocation. Then, by randomly "shuffling" the randomized treatment sequences among all clusters, we generated 1000 permutations from all possible randomized allocations. The permutation distribution of the test statistic was obtained by estimating the treatment effects from the permuted datasets. We rejected the null hypothesis if the test statistic computed from the observed dataset is either smaller than the 2.5th percentile or larger than the 97.5th percentile of the permutation distribution. The whole process was repeated with 500 simulated datasets. The results, presented in Supplementary Table S7, show that permutation test consistently gave inflated type I errors.

## 4. Illustrative example

This section presents a real example to demonstrate the application of the different models considered in this manuscript. OXTEXT-7 was an open cohort SW-CRT involving 11 community mental health teams in the Oxford Health NHS Foundation Trust to assess the effectiveness of a program: "Feeling Well with TrueColours" (an intervention originally used for individuals with bipolar disorder) in producing better health outcomes for participants in their care.[61,62] Teams were assigned at random to receive the intervention throughout 16 months. The Health of the Nation Outcome Scales (HoNOS) total score (continuous) was the primary outcome and was measured monthly, primarily on different people. There were 4595 observations with an average cluster period size of 26. Since the expected churn rate is close to 1 (0.91), for simplicity, we treated this trial as cross-sectional design in the analysis.[63]

Table 6 shows the estimated time-invariant treatment effect, model-based SE and 95% CI for the treatment effect for OXTEXT-7 under all models investigated in the simulations. For models without RVEs, the NE model produced higher SEs. Even though the point estimates slightly differed across the two models, all 95% CIs included zero. Clearly, SEs under EXCH and NE models without RVE were smaller than with any RVE. In this case, it might suggest some degree of random-effect structures misspecification, and hence overly optimistic model-based SE estimates. The RVE results were generally similar across working model specifications for this application. When adding the RVEs, the SE correction for small samples resembled what we saw in our simulations. Not surprisingly, the standard error produced by CR0 was the smallest, but our simulations show that this choice could lead to under-coverage with a small number of clusters, and CR3 is the recommended method. RVEs were generally larger in the NE model (which may be explained by the larger number of random cluster-by-time interactions required by the NE model); however, the CIs were generally similar. When we fitted the RI model, the model-based SE was still much smaller than that under the RVE. This may suggest that the true data generator deviates from the RI model, and based on our simulation results, the RVE, in theory, should be consistent regardless of the random-effects structure specification (assuming the correct fixed-effects specification such that the true treatment effect is time-invariant) and serves as an objective reflection of the uncertainty in the treatment effect estimates. In this example, the DTD-RI model failed to converge. Since the number of periods was larger than the number of clusters, some corrections (e.g., CR1P) would not be available if we treated the period as a categorical variable. One could treat the period as a continuous variable in the analytic model to obtain CR1P, and the impact of this change, particularly for the RVE, is worthy of future investigation.[40,64,65]

One caveat in our illustrative application is that we have assumed that the true treatment effect structure is constant and time-invariant. Although this is the primary focus of our simulation study, this simple treatment effect structure might not hold true for OXTEXT-7, as implied in the reanalysis of Nickless et al.[62] We wish to reiterate that the use of RVEs in linear mixed models may not address the systematic bias that may arise from the incorrect specification of the treatment effect structure (such as specifying a time-invariant treatment effect in the presence of a true exposure-time-specific treatment effect).[66,67] We leave a comprehensive examination of small sample variance corrections under the general time-on-treatment linear mixed model for future research.

## 5. Discussion

### 5.1. Summary of main findings

We examined by simulation whether use of RVEs in linear mixed models can help maintain valid statistical inference under random-effects structure misspecification when the true data are generated under either DTD-RI or NE-RI. Furthermore, we examined the performance of RVEs under five small-sample corrections (available in the R package "clubSandwich") when the number of clusters is limited, which is common in SW-CRTs. We confirmed, as expected with linear mixed models, that misspecification of the correlation structure generally does not introduce bias to the estimated treatment effects. However, random effects misspecification substantially affected the standard errors of estimated treatment effects. This was reflected in large relative errors of model SEs and under-coverage. When the linear mixed model was misspecified, model-based SEs generally underestimated the (true) Monte Carlo SE by 60%. The coverage probability of 95% confidence intervals for the estimated treatment effect dropped to around 50% in misspecified models without RVEs. Similar results have been noticed in previous studies. For example, Kasza and Forbes[24] showed analytically that fitting a simpler correlation structure would lead to an underestimated model-based SE. Our simulation results also confirmed results in Bowden et al.[23] and Voldal et al.[21] that failing to include random treatment effects would lead to an underestimated (model-based) variance of the treatment effect causing lower coverage (or equivalently, inflation of type I error rates). Beyond confirming these findings, our simulations provide compelling evidence that adding RVEs generally improved the validity of statistical inference for a misspecified linear mixed model. This has not been thoroughly discussed in previous literature in the context of SW-CRTs. It is important to notice that even a simple exchangeable model with RVEs (with small-sample correction) has acceptable performance with respect to bias and coverage in SW-CRTs. Adding extra random effects (e.g., using a NE model) did not appreciably improve the model performance in terms of bias and coverage. Regardless of the true data generation mechanism, the performance of different types of RVEs was very similar. Finally, we also wish to point out although we have only considered DTD-RI and NE-RI as two data generators, our findings should be generalizable to other type of linear mixed data generators as the RVE in theory converges to the true variance of the treatment effect estimator under arbitrary random-effects misspecification (assuming the same fixed-effect model).

In the stepped wedge design literature, it is well-known under the GEE framework that the standard RVE could underestimate the variance when the number of clusters is small.[68] Unfortunately, this has not been as widely appreciated for linear mixed models and has led to reservations about adopting linear mixed models with a simple random-effects structure for data analysis. Our study has empirically demonstrated that the RVE can help maintain the validity of inference under a potentially misspecified linear mixed model, and to some extent, provides a path forward in practice when it might be challenging to correctly specify a complicated random-effects structure for SW-CRTs. In our simulations, the number of clusters was the most important factor that affected the performance of RVEs. The relative error of the model SE produced by using standard RVE (CR0) without small sample correction could still be as high as -22%, and the coverage probability could be as low as 83%. In fact, with fewer than 32 clusters, we found that small-sample corrections were essential. Among five types of small-sample correction methods (CR1, CR1P, CR1S, CR2, and CR3), when a trial has a relatively large number of clusters (e.g., 32), all types of small-sample corrections had similar performance. With small number of clusters, CR2 performed reasonably well and was clearly better than the others but exhibited some mild under-coverage. Using CR3 generally led to similar performance to the true model, small relative errors of model-based SEs, and coverage probabilities close to 95%. In our simulations, we found that models with CR3 (DoF = $I - 2$) performed adequately but tended to have conservative coverage probabilities when the number of clusters was only 8. This observation is consistent with prior findings in the GEE literature.[54] The use of Satterthwaite DoF seems to be more prevalent in practice for linear mixed models, but we found that it generally led to coverage probabilities lower than the nominal level. Although the DoF correction $I - 2$ has previously been implemented in a GEE framework with RVE, we have shown that it is equally applicable to linear mixed models with RVEs, and indeed we have empirically demonstrated its appropriate performance in our simulations with a small number of clusters.

### 5.2. Recommendations and limitations

To the best of our knowledge, our study is the first to comprehensively examine the performance of misspecified random-effects structures in linear mixed models with RVEs in the context of SW-CRTs. When the true correlation structure is unknown, we found that a random intercept model with standard RVE (CR0) is usually sufficient to achieve valid inference with at least 32 clusters. A small-sample correction (e.g., CR3) is highly recommended for cases with fewer than 32 clusters. With the number of clusters minus two DoF correction, CR3 achieved appropriate 95% CI coverage with as few as 8 clusters, at the expense of a slight overestimation of SEs. Putting our results in the context of the SW-CRT literature, we notice that our findings are slightly different from those obtained from previous simulations of SW-CRTs with GEE estimators. To give a few examples, Li[47] and Li et al.[41] have found that the Kauermann and Carroll variance estimator (the equivalence of CR2 under GEE inference), coupled with the DoF $I - 2$, can maintain nominal type I error rates for GEE analysis of small SW-CRTs; Ford and Westgate[54] found that an average of the Kauermann and Carroll SE and Mancl and DeRouen SE (the equivalence of CR3 under GEE inference) estimators, coupled with the DoF $I - 2$, can maintain nominal type I error rates for GEE analysis of small SW-CRTs. However, apart from the fact that these prior studies considered a different estimator than ours, they have mostly investigated the performance of RVE under correct specification or approximately correct specification of the within-cluster correlation structure (as opposed to misspecified random-effects structures). In our simulations, CR3 coupled with a $t(I - 2)$ reference distribution is recommended as it maintained the best small-sample inference for misspecified linear mixed models in SW-CRTs.

Historically, GEE is known to be more robust than linear mixed models but usually requires larger numbers of clusters. Our findings empirically demonstrate that linear mixed models may be equally — if not more — competitive than GEE in providing robust analysis of small SW-CRTs. In fact, for continuous outcomes, the performance of GEE and linear mixed models is very similar when the marginal model (fixed effects) is correctly specified: their differences lie in the estimation of nuisance parameters such as the variance components and the correlation parameters. A potential advantage of linear mixed models is that it is generally easier to specify and estimate more complicated random-effects structures, whereas substantial conceptual difficulty and computational burden may be introduced to the formulation and estimation of an equally complex working correlation structure with multiple parameters in GEE.[41] In the analysis of SW-CRTs, there is a potential efficiency advantage in specifying a more complicated correlation structure, especially when the fitted model with RVE is closer to the true model (as also shown in our simulation Supplementary Table S5). For example, when the true model is NE+RI, the nested exchangeable model with RVE has slightly higher power than the exchangeable model with RVE. To specify the former structure beyond simple exchangeability, the linear mixed model is arguably much more accessible and may have a computational advantage over GEE when the cluster size is large. On the other hand, GEE inference may enjoy other advantages such as directly reporting the correlation parameters on the natural scale of the outcomes and may be particularly preferred for the analysis of binary and count outcomes in SW-CRTs with a non-identity link function.[39–41,56] In a pair of recent systematic reviews of 160 published SW-CRTs by Nevins et al.,[26,48] 70% (112/160) adopted mixed-effects model in the primary analysis of the primary outcome, likely due to software accessibility and flexibility in specifying the random-effects structure. However, although not examined in the previous systematic review, our experience is that RVE is not frequently used in practice, possibly to guard against anti-conservative inference under random-effects misspecification. Therefore, an important contribution of our study is promoting awareness and empirically demonstrating the use of RVE as a reliable tool for robust inference in misspecified linear mixed models in SW-CRTs.

Previous studies have indicated that design-based inference, such as the permutation test, is a powerful tool to provide robust inference under model misspecification. With a limited number of clusters, one clear advantage of the permutation test is that the inferences are exact such that the type I error rate can often be controlled at the nominal level without the need for small sample corrections. For example, Wang and DeGruttola[69] and Ji et al.[70] have considered testing the null hypothesis based on identifying the permutational distribution of the Wald statistic estimated from the simple exchangeable and the nested exchangeable random-effects model. In the same spirit of the use of RVE in our article, they both demonstrated that the specification of the random-effects has little impact on the Type I error rate but only affected the power of the permutation test. Alternative test statistics with fewer modeling assumptions in SW-CRTs have subsequently been discussed by Thompson et al.,[71] Kennedy-Shaffer et al.,[72] Hughes et al,[59] under the principle of conditional exchangeability (which often serves as the basis for permutation inference). Zhang and Zhao[73] further carefully distinguished between the permutation test and the randomization test, and illustrated the practical importance of this distinction with a SW-CRT example. More recently, Ren et al.[60] have found that the permutation test may not protect against all types of random effect misspecification. Specifically, they found that when the true model includes a random intervention effect, but the fitted model did not account for that, the Type I error rate of the permutation test based on a Wald statistic was inflated. This matches our additional simulation results and is mainly because the treatment sequence affects both the mean and variance structures, leading to violations of conditional exchangeability even under the null based on the working model with a random intervention effect. In this scenario, however, the use of RVEs based on a misspecified linear mixed model can successfully preserve the validity of inference, and therefore could be preferred. Finally, it is worth noting that the construction of the permutation-based confidence intervals can require substantial computation from inverting the permutation test. As an improvement, Rabideau and Wang[74] recently studied more efficient search algorithms for obtaining permutation confidence intervals for SW-CRTs. In comparison, the construction of the confidence interval based on a RVE is simple and essentially instantaneously computed. We provide in Table 7 of our revision a concise summary of the potential advantages and caveats of using RVEs in mixed-effects models, RVEs in GEE, and permutation tests to provide robust inference in SW-CRTs.

In practice, while we still maintain the usual standpoint of carefully specifying the random-effects structure in SW-CRTs (for reasons such that we could obtain intra-cluster correlation coefficients that are essential for future trial planning,[14] or that we might be able to obtain a more efficient treatment effect estimator when the specified random-effects structure approximates the truth), RVEs are recommended when a more complex model cannot be fit or fails to converge, or researchers are strongly concerned about model misspecification. It is also recommended to fit a model with a RVE as a sensitivity analysis even if researchers have a strong belief about the true correlation structure — in such a case one would expect the results to be similar between the model-based variance estimator and the RVE if the number of clusters is large enough. We also conjecture that, because treatment effects in other types of multiple-period cluster randomized trial designs are often estimated via linear mixed models, our results can be generalized to such other types of CRTs (such as longitudinal parallel-arm CRTs and cluster randomized crossover trials).

Non-convergence is another practical issue when the fitted linear mixed model involves a complicated random-effects structure. We observed more than 30% non-convergence rate for the DTD-RI model when the trial is small. In practice, the DTD-RI model is still encouraged if the investigators believe it reflects their beliefs about the true random effect correlation structures. In case of convergence failure, investigators could change the default optimizer or follow the troubleshooting documentation included in the "glmmTmb" package. SAS and STAT users may follow some strategies denoted in an instruction document by Kiernan.[75]

Our study has several limitations. First, although we have demonstrated that appropriate inference for treatment effects can be obtained from misspecified random-effects structures in linear mixed models by using the RVE, we wish to point out that the estimates of the random-effects variance components from misspecified models will no longer be valid. If the goal is to obtain valid ICC estimates (e.g., to inform future sample size calculations), then the interpretation of the variance component and ICC estimates from a misspecified working model can become challenging.[21,23,76] An exception is discussed by Kasza et al,[76] where one may obtain decay parameter estimates from variance component estimates under a misspecified exchangeable random-effects model. Second, we only considered continuous outcomes as RVEs for non-Gaussian outcomes are currently not available across all software packages. For example, the "clubSandwich" R package does not support non-Gaussian outcomes. Even though SAS GLIMMIX allows for RVE, the estimation routine for a generalized linear mixed model proceeds by linearization and should be considered as pseudo-GEE. However, a more conceptual challenge with using misspecified generalized linear mixed models with binary outcomes, and particularly with a logit link, is that the treatment effect coefficient may carry a different interpretation when the random-effects structure is misspecified, due to non-collapsibility.[77] Future studies should be carried out to reconcile the challenges in using misspecified generalized linear mixed models in SW-CRTs. Despite of all these complexities, the linear mixed models we considered can still be used to estimate the treatment effect on the risk difference scale for binary outcomes, in which case the RVE becomes essential as it can simultaneously account for a misspecified variance function and potentially misspecified random-effects structure. Third, when considering the models used to generate the data, we assumed the DTD-RI was the most complex model. In fact, more complex correlation structures, such as Toeplitz and unstructured, have been discussed previously.[31] Although RVE is in theory valid under more complex linear mixed data generators, its performance under such scenarios was not examined in this study. Fourth, in this study, we still found that when the number of clusters is small (e.g., 8), RVE with small-sample correction tended to have slightly inflated Type I error rates. Under the GEE framework, alternative small-sample corrected RVEs (such as that due to Fay and Graubard[78]) have been studied, but under relatively restrictive data generating processes.[27,79–82] However, such bias-reduced estimator has not been widely implemented for linear mixed models and its performance with a potentially misspecified mixed model is unclear. Currently, the Fay and Graubard[78] variance estimator for linear mixed models is only available in SAS PROC

GLIMMIX (which fits mixed models via a GEE-like routine rather than a maximum likelihood routine). Future studies may consider evaluating this additional RVE for linear mixed models fitted in SAS PROC GLIMMIX. Lastly, we only discussed cross-sectional designs. The cohort design requires additional random effects for repeated measures on each individual. We expect that our results can be extended to cohort designs, but future studies should investigate the performance of each RVE when additional random effects are specified for repeated measurements.

# Acknowledgement

This research was enabled in part by support provided by Compute Ontario (computeontario.ca) and the Digital Research Alliance of Canada (alliancecan.ca)

# Declaration of conflicting interests

The Authors declare that there is no conflict of interest.

# Funding

M.T. and F.L. are supported by the National Institute of Aging (NIA) of the National Institutes of Health (NIH) under Award Number U54AG063546, which funds NIA Imbedded Pragmatic Alzheimer's Disease and AD-Related Dementias Clinical Trials Collaboratory (NIA IMPACT Collaboratory). Research in this article was also partially supported by a Patient-Centered Outcomes Research Institute Award® (PCORI® Award ME-2022C2-27676, to F.L.). Y.O. is funded by the Canadian Institutes of Health Research Health System Impact Postdoc Fellowship. The statements presented in this article are solely the responsibility of the authors and do not necessarily represent the official views of NIH, PCORI®, or its Board of Governors or Methodology Committee.

# Data availability statement

Data sharing is not applicable to this article as no new data were created or analyzed in this study. Some code used for simulation is available at https://github.com/douyangyd/swcrt_rve

**Table 1 Summary of robust variance estimators used in our simulations**

| Robust variance estimator | Description |
|---|---|
| CR0 | Standard robust variance estimator |
| CR1 | Simple degrees of freedom (DF) correction of CR0 in the form $I/(I-1)$ |
| CR1P | Simple degrees of freedom (DF) correction of CR0 in the form $I/(I-P)$ where is $P$ is the number of covariate parameters estimated in the model |
| CR1S | Simple degrees of freedom (DF) correction of CR0 in the form $(I(N-1))/[(I-1)(N-P)]$ where is $N$ is the number of observations |
| CR2 | Correction based on "bias-reduced linearization" |
| CR3 | Closely approximates the leave-one-cluster-out jackknife resampling estimator |

**Table 2 Summary of within-period and between-period intracluster correlation coefficients under data generating models**

| Data generating model | Parameters | Under control condition | Under intervention condition |
|---|---|---|---|
| **Random intervention with nested exchangeable** | Within-period ICC | $\dfrac{\tau_\alpha^2 + \tau_\gamma^2}{\tau_\alpha^2 + \tau_\gamma^2 + \tau_\varepsilon^2}$ | $\dfrac{\tau_\alpha^2 + \tau_\gamma^2 + \tau_\nu^2}{\tau_\alpha^2 + \tau_\gamma^2 + \tau_\nu^2 + \tau_\varepsilon^2}$ |
| | Between-period ICC | $\dfrac{\tau_\alpha^2}{\tau_\alpha^2 + \tau_\gamma^2 + \tau_\varepsilon^2}$ | $\dfrac{\tau_\alpha^2 + \tau_\nu^2}{\tau_\alpha^2 + \tau_\gamma^2 + \tau_\nu^2 + \tau_\varepsilon^2}$ |
| | CAC | $\dfrac{\tau_\alpha^2}{\tau_\alpha^2 + \tau_\gamma^2}$ | $\dfrac{\tau_\alpha^2 + \tau_\nu^2}{\tau_\alpha^2 + \tau_\gamma^2 + \tau_\nu^2}$ |
| **Random intervention with discrete time decay (assuming an AR(1) $\widetilde{\mathbf{Z}}$ matrix)** | Within-period ICC | $\dfrac{\tau_\gamma^2}{\tau_\gamma^2 + \tau_\varepsilon^2}$ | $\dfrac{\tau_\gamma^2 + \tau_\nu^2}{\tau_\gamma^2 + \tau_\nu^2 + \tau_\varepsilon^2}$ |
| | CAC (between period $j$ and $l$, where $j \neq l$) | $r^{|j-l|}$ | $\dfrac{\tau_\gamma^2 r^{|j-l|} + \tau_\nu^2}{\tau_\gamma^2 + \tau_\nu^2}$ |
| | Between-period ICC (between period $j$ and $l$, where $k \neq l$) | $\dfrac{\tau_\gamma^2 r^{|j-l|}}{\tau_\gamma^2 + \tau_\varepsilon^2}$ | $\dfrac{\tau_\gamma^2 r^{|j-l|} + \tau_\nu^2}{\tau_\gamma^2 + \tau_\nu^2 + \tau_\varepsilon^2}$ |

ICC: intracluster correlation coefficient; CAC: cluster autocorrelation; $r_{jl}$: the decay parameter (CAC) between period $i$ and period $j$ assuming a general $\widetilde{\mathbf{Z}}$ matrix in the DTD-RI model. $\rho$: the decay parameter (CAC) in adjacent periods assuming an AR(1) $\widetilde{\mathbf{Z}}$ matrix in the DTD-RI model.

**Table 3 Range of trial configuration and simulation parameter values under random intervention with nested exchangeable correlation structure model**

| Parameter | Values |
|---|---|
| **Number of clusters (I)** | 8, 16, 32 |
| **Number of sequences (S)** | 4, 8 |
| **Cluster-period sizes (K)** | 10, 100 |
| **Effect size** | 0 |
| **Within-period ICC in control and intervention group** | (0.01, 0.05), (0.05, 0.10), (0.05, 0.15) |
| **CAC** | 0.8 |
| **Standard deviation of individual error term** | 1 |

ICC: intracluster correlation coefficient; CAC: cluster autocorrelation

**Table 4 Summary of performance measures**

| Performance measures | Definition | Estimates |
|---|---|---|
| **Bias** | $E[\hat{\theta}] - \theta$ | $\frac{1}{n_{sim}} \sum_{i=1}^{n_{sim}} \widehat{\theta_i} - \theta$ |
| **Coverage** | $P(\widehat{\theta_{low}} < \theta < \widehat{\theta_{upp}})$ | $\frac{1}{n_{sim}} \sum_{i=1}^{n_{sim}} I(\widehat{\theta_{low}} < \theta_i < \widehat{\theta_{upp}})$ |
| **Average Estimated SE** | $E\left[\sqrt{\widehat{Var}(\hat{\theta})}\right]$ | $\frac{1}{n_{sim}} \sum_{i=1}^{n_{sim}} \sqrt{\widehat{Var}(\widehat{\theta_i})}$ |
| **Empirical SE** | $\sqrt{Var(\hat{\theta})}$ | $\sqrt{\frac{1}{n_{sim}} \sum_{i=1}^{n_{sim}} (\widehat{\theta_i} - \bar{\theta})^2}$ |
| **Percentage error of average estimated SE to empirical SE** | $100 \left( \frac{E\left[\sqrt{\widehat{Var}(\hat{\theta})}\right]}{\sqrt{Var(\hat{\theta})}} - 1 \right)$ | $100 \left( \frac{\frac{1}{n_{sim}} \sum_{i=1}^{n_{sim}} \sqrt{\widehat{Var}(\widehat{\theta_i})}}{\sqrt{\frac{1}{n_{sim}} \sum_{i=1}^{n_{sim}} (\widehat{\theta_i} - \bar{\theta})^2}} - 1 \right)$ |

SE: standard error

**Table 5 Percentage of non-convergence for fitting the true model across 2,000 simulations**

| I | S | WPICC (Control, Intervention) | K | NE-RI* | NE-RI** | DTD-RI |
|---|---|---|---|---|---|---|
| 8 | 4 | (0.01, 0.05) | 10 | 0.3% | 0.0% | 39.4% |
| 8 | 4 | (0.01, 0.05) | 100 | 3.2% | 0.2% | 23.1% |
| 8 | 4 | (0.05, 0.10) | 10 | 0.9% | 0.2% | 30.8% |
| 8 | 4 | (0.05, 0.10) | 100 | 1.9% | 0.1% | 1.7% |
| 8 | 4 | (0.05, 0.15) | 10 | 1.0% | 0.0% | 30.4% |
| 8 | 4 | (0.05, 0.15) | 100 | 1.4% | 0.2% | 1.7% |
| 8 | 8 | (0.01, 0.05) | 10 | 1.2% | 0.2% | 27.0% |
| 8 | 8 | (0.01, 0.05) | 100 | 4.5% | 0.1% | 7.9% |
| 8 | 8 | (0.05, 0.10) | 10 | 1.8% | 0.2% | 12.8% |
| 8 | 8 | (0.05, 0.10) | 100 | 3.7% | 0.0% | 0.1% |
| 8 | 8 | (0.05, 0.15) | 10 | 1.2% | 0.1% | 13.5% |
| 8 | 8 | (0.05, 0.15) | 100 | 3.4% | 0.0% | 0.0% |
| 16 | 4 | (0.01, 0.05) | 10 | 1.2% | 0.0% | 34.6% |
| 16 | 4 | (0.01, 0.05) | 100 | 4.8% | 0.1% | 12.5% |
| 16 | 4 | (0.05, 0.10) | 10 | 1.3% | 0.2% | 20.4% |
| 16 | 4 | (0.05, 0.10) | 100 | 2.6% | 0.0% | 0.0% |
| 16 | 4 | (0.05, 0.15) | 10 | 1.1% | 0.0% | 21.1% |
| 16 | 4 | (0.05, 0.15) | 100 | 2.5% | 0.0% | 0.0% |
| 16 | 8 | (0.01, 0.05) | 10 | 1.6% | 0.1% | 21.0% |
| 16 | 8 | (0.01, 0.05) | 100 | 8.2% | 0.1% | 0.9% |
| 16 | 8 | (0.05, 0.10) | 10 | 2.6% | 0.1% | 5.4% |
| 16 | 8 | (0.05, 0.10) | 100 | 6.1% | 0.0% | 0.0% |
| 16 | 8 | (0.05, 0.15) | 10 | 1.6% | 0.1% | 5.7% |
| 16 | 8 | (0.05, 0.15) | 100 | 6.6% | 0.0% | 0.0% |
| 32 | 4 | (0.01, 0.05) | 10 | 1.8% | 0.1% | 31.9% |
| 32 | 4 | (0.01, 0.05) | 100 | 7.7% | 0.1% | 4.6% |
| 32 | 4 | (0.05, 0.10) | 10 | 1.4% | 0.2% | 10.9% |
| 32 | 4 | (0.05, 0.10) | 100 | 4.9% | 0.0% | 0.0% |
| 32 | 4 | (0.05, 0.15) | 10 | 1.1% | 0.2% | 12.3% |
| 32 | 4 | (0.05, 0.15) | 100 | 5.9% | 0.0% | 0.0% |
| 32 | 8 | (0.01, 0.05) | 10 | 2.7% | 0.1% | 17.9% |
| 32 | 8 | (0.01, 0.05) | 100 | 14.5% | 0.0% | 0.1% |
| 32 | 8 | (0.05, 0.10) | 10 | 3.4% | 0.0% | 1.3% |
| 32 | 8 | (0.05, 0.10) | 100 | 12.7% | 0.0% | 0.0% |
| 32 | 8 | (0.05, 0.15) | 10 | 2.7% | 0.1% | 1.4% |
| 32 | 8 | (0.05, 0.15) | 100 | 11.3% | 0.0% | 0.0% |

I: Number of clusters; S: Number of sequences, M: Cluster-period size
*Linear mixed effect model with optimizer "nloptwarp"
** Linear mixed effect model with optimizer "bobyqa"

**Table 6 Example OXTEXT-7 trial analyzed with each method investigated**

| Model | Estimated time-invariant treatment effects | Estimated SE | 95% confidence interval |
|---|---|---|---|
| **Exchangeable**** | 0.066 | 0.402 | (-0.843, 0.975) |
| **Exchangeable + CR0** | | 0.651 | (-1.407, 1.539) |
| **Exchangeable + CR1** | | 0.683 | (-1.479, 1.611) |
| **Exchangeable + CR1P*** | | - | - |
| **Exchangeable + CR1S** | | 0.684 | (-1.481, 1.613) |
| **Exchangeable + CR2** | | 0.706 | (-1.531, 1.663) |
| **Exchangeable + CR3** | | 0.767 | (-1.669, 1.801) |
| | | | |
| **Nested Exchangeable**** | 0.238 | 0.487 | (-0.864, 1.340) |
| **Nested Exchangeable + CR0** | | 0.711 | (-1.370, 1.846) |
| **Nested Exchangeable + CR1** | | 0.746 | (-1.450, 1.926) |
| **Nested Exchangeable + CR1P*** | | - | - |
| **Nested Exchangeable + CR1S** | | 0.747 | (-1.452, 1.928) |
| **Nested Exchangeable + CR2** | | 0.762 | (-1.486, 1.962) |
| **Nested Exchangeable + CR3** | | 0.820 | (-1.617, 2.093) |
| | | | |
| **NE-RI**** | 0.322 | 0.504 | (-0.818, 1.462) |
| **DTD-RI** | Did Not Converge | | |

NE-RI: nested exchangeable + random intervention model
DTD-RI: discrete decay time + random intervention model
*CR1P is not available as the number of parameters in the model is greater than number of clusters
**For model without RVE, the model-based SE was presented

**Table 7 A concise comparison of methods regarding their features for robust analyses of SW-CRTs under model misspecification.**

| Methods | Potential advantages | Caveats |
|---|---|---|
| **Mixed-effects models with RVE** | <ul><li>It is convenient to specify more complicated random-effects structures to account for multiple sources of intracluster correlations.</li><li>The implementation is available in standard software.</li></ul> | <ul><li>For logistic link function and binary outcomes, the interpretation of the fixed-effects parameter may change according to different specifications of the random-effects structure within a generalized linear mixed model.</li><li>The use of RVE requires small sample correction with a small number of clusters.</li></ul> |
| **GEE models with RVE** | <ul><li>The estimation of the marginal mean structure and the correlation structure is based on separate estimating equations, and misspecification of the correlation structure does not affect the interpretation of the marginal mean parameters, regardless of the type of the outcome and link function.</li><li>GEE with a simple exchangeable correlation structure is implemented in standard software</li></ul> | <ul><li>Recently developed software for implementing GEE with more complicated correlation structure in SW-CRTs may be limited to certain two-parameter of three-parameter correlation structures.</li><li>When specifying a more complicated correlation structure, it is likely that one needs to provide the design matrix for the correlation estimating equations, which requires careful coding.</li><li>The computation is often intensive for with a complex correlation structure when the cluster size becomes large.</li><li>The use of RVE requires small sample correction with a small number of clusters.</li></ul> |
| **Permutation inference** | <ul><li>Under conditional exchangeability, the inference is exact and does not require small sample correction even when there is a small number of clusters.</li></ul> | <ul><li>Estimating the permutation distribution of the test statistic may require more computation due to repeatedly fitting multilevel models under permuted treatment assignments.</li><li>Estimation of the permutation confidence interval is not trivial and requires substantial computation from a search algorithm.</li></ul> |

**Figure 1** Illustration of a stepped wedge design, with five sequences, six periods

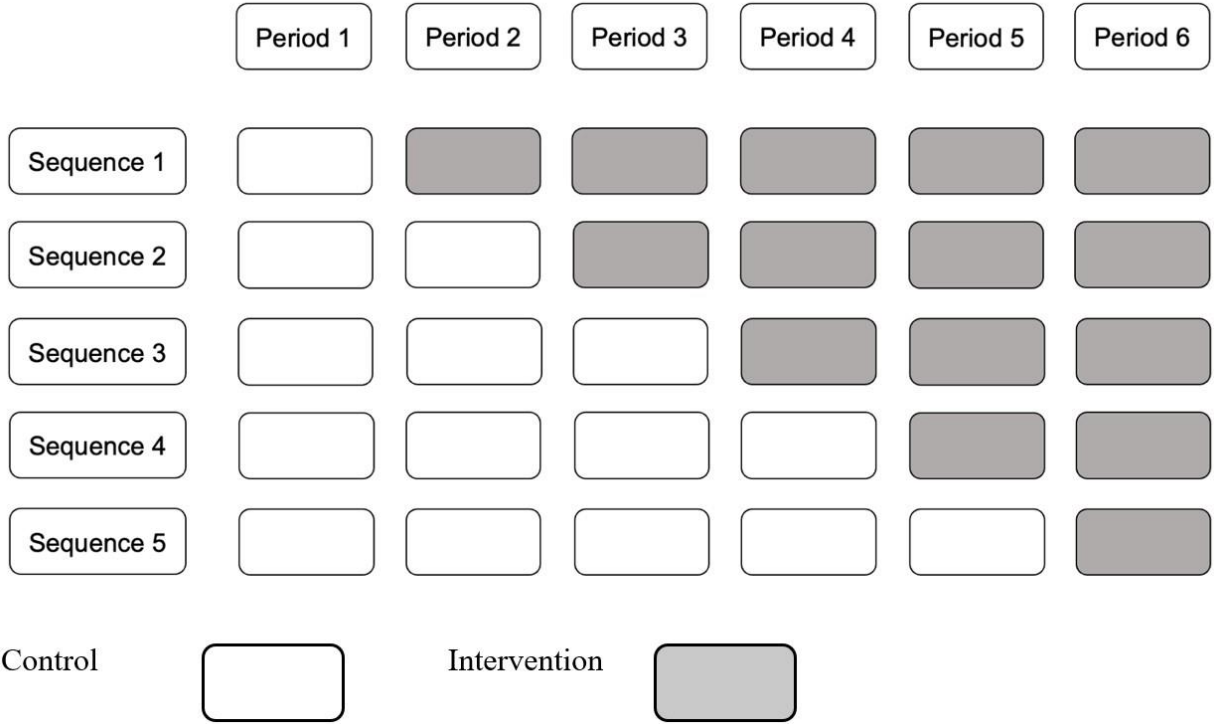

**Figure 2 Coverage probability of 95% confidence intervals for the treatment effect under both the true model and misspecified models with or without RVE when data were generated under DTD-Rl**

CR0 is the standard robust variance estimator, and CR3 closely approximates the leave-one-cluster-out jackknife resampling estimator. The three dashed lines represent the 95% $\pm$ 2*MCSE.

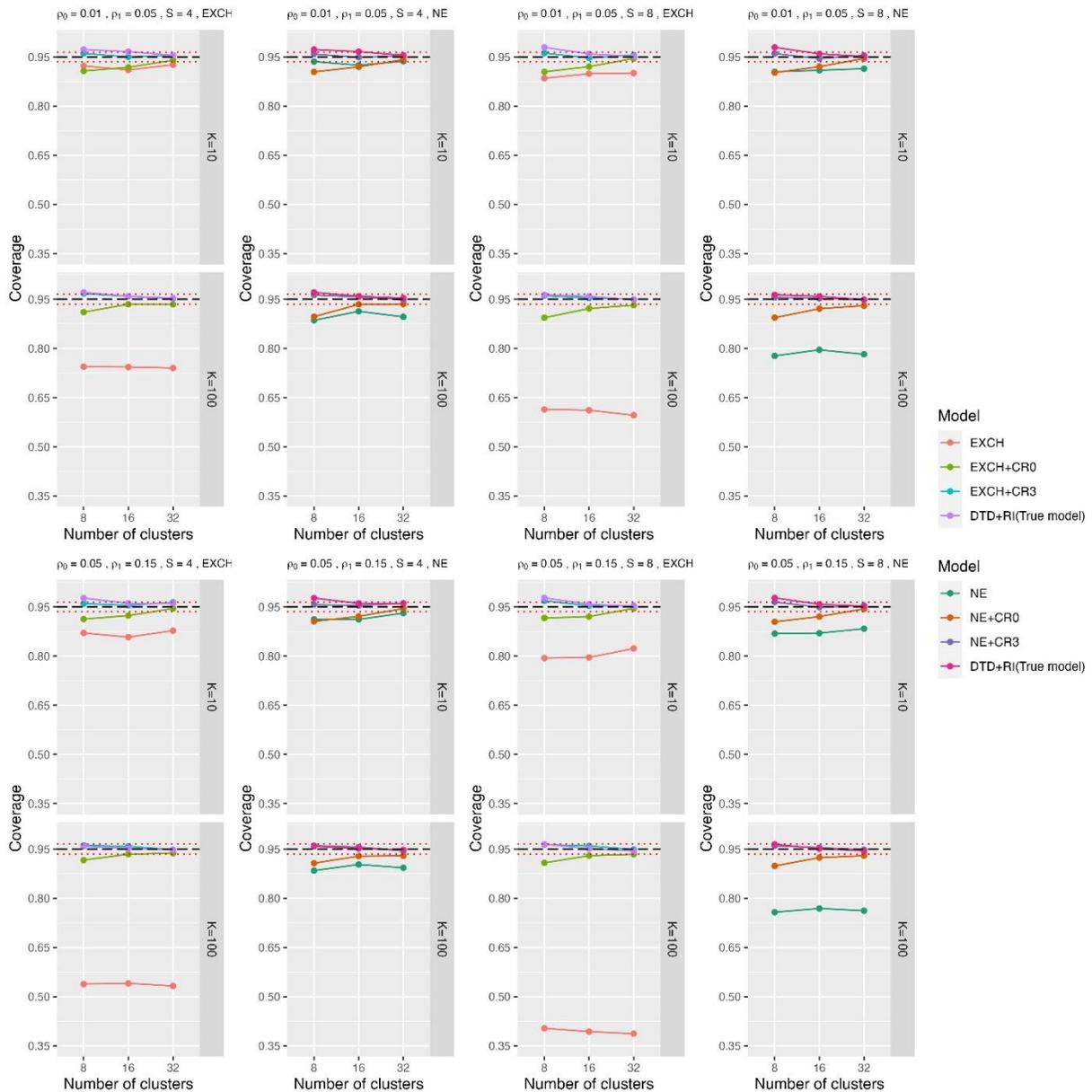

**Figure 3 Coverage probability of 95% confidence intervals for the treatment effect under both the true model and misspecified models with or without RVE when data were generated under NE-Rl**

CR0 is the standard robust variance estimator, and CR3 closely approximates the leave-one-cluster-out jackknife resampling estimator. The three dashed lines represent the 95% $\pm$ 2*MCSE.

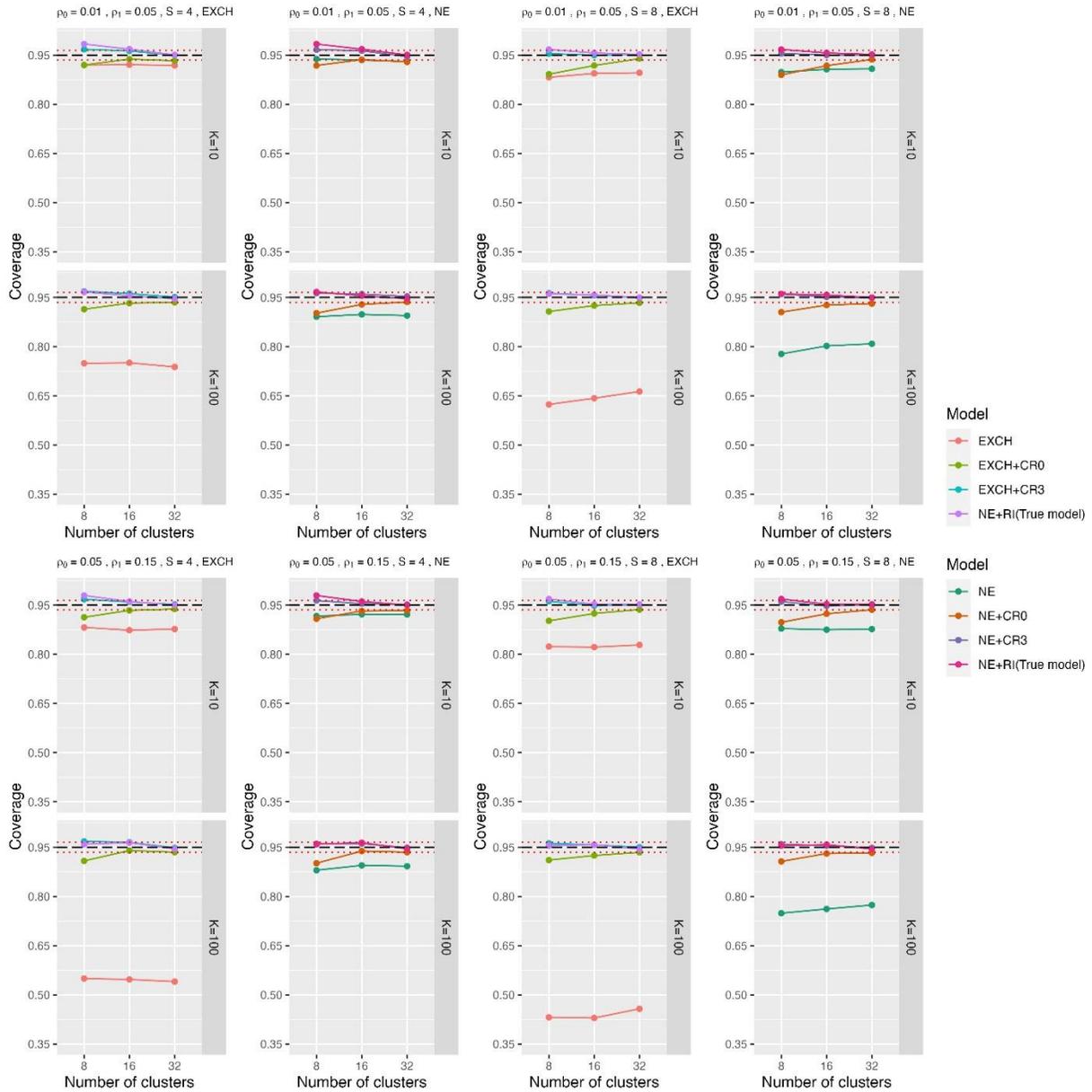

# Table S1 Estimated Bias of treatment effects

| | | True model - NE + Random intervention | | | | |
|---|---|---|---|---|---|---|
| I | S | WPICC (Control, Intervention) | K | EXCH | NE | TRUE |
| 8 | 4 | (0.01, 0.05) | 10 | -0.002 | -0.003 | -0.003 |
| 8 | 4 | (0.01, 0.05) | 100 | 0.000 | 0.001 | 0.001 |
| 8 | 4 | (0.05, 0.10) | 10 | -0.001 | -0.001 | -0.002 |
| 8 | 4 | (0.05, 0.10) | 100 | 0.001 | 0.002 | 0.002 |
| 8 | 4 | (0.05, 0.15) | 10 | -0.001 | 0.000 | -0.002 |
| 8 | 4 | (0.05, 0.15) | 100 | 0.002 | 0.002 | 0.003 |
| 8 | 8 | (0.01, 0.05) | 10 | 0.002 | 0.002 | 0.002 |
| 8 | 8 | (0.01, 0.05) | 100 | 0.002 | 0.002 | 0.003 |
| 8 | 8 | (0.05, 0.10) | 10 | 0.003 | 0.003 | 0.004 |
| 8 | 8 | (0.05, 0.10) | 100 | 0.001 | 0.001 | 0.003 |
| 8 | 8 | (0.05, 0.15) | 10 | 0.004 | 0.004 | 0.004 |
| 8 | 8 | (0.05, 0.15) | 100 | 0.002 | 0.002 | 0.004 |
| 16 | 4 | (0.01, 0.05) | 10 | 0.000 | 0.000 | 0.000 |
| 16 | 4 | (0.01, 0.05) | 100 | 0.000 | 0.000 | -0.001 |
| 16 | 4 | (0.05, 0.10) | 10 | 0.002 | 0.002 | 0.002 |
| 16 | 4 | (0.05, 0.10) | 100 | 0.006 | 0.006 | 0.006 |
| 16 | 4 | (0.05, 0.15) | 10 | 0.004 | 0.005 | 0.004 |
| 16 | 4 | (0.05, 0.15) | 100 | -0.005 | -0.005 | -0.004 |
| 16 | 8 | (0.01, 0.05) | 10 | 0.001 | 0.001 | 0.001 |
| 16 | 8 | (0.01, 0.05) | 100 | -0.001 | -0.001 | -0.001 |
| 16 | 8 | (0.05, 0.10) | 10 | 0.001 | 0.001 | 0.001 |
| 16 | 8 | (0.05, 0.10) | 100 | -0.002 | -0.002 | -0.002 |
| 16 | 8 | (0.05, 0.15) | 10 | 0.000 | 0.000 | 0.000 |
| 16 | 8 | (0.05, 0.15) | 100 | -0.003 | -0.002 | -0.003 |
| 32 | 4 | (0.01, 0.05) | 10 | -0.001 | -0.001 | -0.002 |
| 32 | 4 | (0.01, 0.05) | 100 | 0.000 | 0.000 | 0.000 |
| 32 | 4 | (0.05, 0.10) | 10 | -0.002 | -0.002 | -0.002 |
| 32 | 4 | (0.05, 0.10) | 100 | -0.001 | 0.000 | 0.000 |
| 32 | 4 | (0.05, 0.15) | 10 | -0.002 | -0.002 | -0.002 |
| 32 | 4 | (0.05, 0.15) | 100 | -0.001 | 0.000 | -0.001 |
| 32 | 8 | (0.01, 0.05) | 10 | 0.000 | -0.001 | 0.000 |
| 32 | 8 | (0.01, 0.05) | 100 | -0.001 | -0.001 | -0.001 |
| 32 | 8 | (0.05, 0.10) | 10 | 0.003 | 0.003 | 0.003 |
| 32 | 8 | (0.05, 0.10) | 100 | -0.001 | 0.000 | 0.000 |
| 32 | 8 | (0.05, 0.15) | 10 | -0.001 | -0.001 | -0.001 |
| 32 | 8 | (0.05, 0.15) | 100 | -0.002 | -0.002 | -0.002 |
| | | True model - DTD + Random intervention | | | | |
| I | S | WPICC (Control, Intervention) | K | EXCH | NE | TRUE |
| 8 | 4 | (0.01, 0.05) | 10 | -0.003 | -0.002 | -0.006 |
| 8 | 4 | (0.01, 0.05) | 100 | 0.000 | 0.000 | -0.001 |
| 8 | 4 | (0.05, 0.10) | 10 | -0.002 | -0.003 | -0.005 |
| 8 | 4 | (0.05, 0.10) | 100 | 0.000 | 0.000 | -0.001 |
| 8 | 4 | (0.05, 0.15) | 10 | -0.002 | -0.002 | -0.004 |
| 8 | 4 | (0.05, 0.15) | 100 | 0.000 | 0.000 | -0.002 |
| 8  | 8 | (0.01, 0.05) | 10  | -0.001 | -0.001 | -0.003 |
| 8  | 8 | (0.01, 0.05) | 100 | 0.001  | 0.002  | 0.001  |
| 8  | 8 | (0.05, 0.10) | 10  | 0.000  | 0.000  | -0.002 |
| 8  | 8 | (0.05, 0.10) | 100 | 0.001  | 0.002  | 0.001  |
| 8  | 8 | (0.05, 0.15) | 10  | 0.000  | 0.000  | -0.003 |
| 8  | 8 | (0.05, 0.15) | 100 | 0.002  | 0.002  | 0.002  |
| 16 | 4 | (0.01, 0.05) | 10  | 0.001  | 0.001  | 0.000  |
| 16 | 4 | (0.01, 0.05) | 100 | 0.001  | 0.001  | 0.001  |
| 16 | 4 | (0.05, 0.10) | 10  | 0.000  | 0.000  | -0.001 |
| 16 | 4 | (0.05, 0.10) | 100 | 0.001  | 0.001  | 0.001  |
| 16 | 4 | (0.05, 0.15) | 10  | 0.001  | 0.002  | 0.001  |
| 16 | 4 | (0.05, 0.15) | 100 | 0.002  | 0.001  | 0.001  |
| 16 | 8 | (0.01, 0.05) | 10  | 0.001  | 0.000  | 0.000  |
| 16 | 8 | (0.01, 0.05) | 100 | 0.001  | 0.001  | 0.001  |
| 16 | 8 | (0.05, 0.10) | 10  | 0.001  | 0.001  | 0.002  |
| 16 | 8 | (0.05, 0.10) | 100 | 0.001  | 0.001  | 0.001  |
| 16 | 8 | (0.05, 0.15) | 10  | 0.001  | 0.001  | 0.002  |
| 16 | 8 | (0.05, 0.15) | 100 | 0.001  | 0.001  | 0.001  |
| 32 | 4 | (0.01, 0.05) | 10  | 0.002  | 0.002  | 0.000  |
| 32 | 4 | (0.01, 0.05) | 100 | 0.001  | 0.001  | 0.001  |
| 32 | 4 | (0.05, 0.10) | 10  | 0.001  | 0.001  | 0.001  |
| 32 | 4 | (0.05, 0.10) | 100 | -0.001 | 0.000  | -0.001 |
| 32 | 4 | (0.05, 0.15) | 10  | 0.000  | 0.001  | 0.001  |
| 32 | 4 | (0.05, 0.15) | 100 | -0.001 | 0.000  | -0.001 |
| 32 | 8 | (0.01, 0.05) | 10  | 0.000  | 0.000  | -0.001 |
| 32 | 8 | (0.01, 0.05) | 100 | -0.001 | -0.001 | -0.001 |
| 32 | 8 | (0.05, 0.10) | 10  | 0.000  | 0.000  | -0.001 |
| 32 | 8 | (0.05, 0.10) | 100 | -0.003 | -0.003 | -0.002 |
| 32 | 8 | (0.05, 0.15) | 10  | 0.000  | 0.000  | -0.001 |
| 32 | 8 | (0.05, 0.15) | 100 | -0.002 | -0.003 | -0.002 |

**Table S2 Coverage probability of 95% confidence interval (Degree of Freedom: Number of clusters minus two)**
Colored cells: Coverage Probabilities that are within 2*MCSE

| True model = NE + Random intervention | | | | EXCH | | | | | | | NE | | | | | | | True model |
|---|---|---|---|---|---|---|---|---|---|---|---|---|---|---|---|---|---|---|
| I | S | WPICC (Control, Intervention) | K | Standard | CR0 | CR1 | CR1P | CR1S | CR2 | CR3 | Standard | CR0 | CR1 | CR1P | CR1S | CR2 | CR3 | |
| 8 | 4 | (0.01, 0.05) | 10 | 92.0% | 92.0% | 93.5% | 99.5% | 93.6% | 95.0% | 96.8% | 93.9% | 91.9% | 93.4% | 99.4% | 93.5% | 94.9% | 96.7% | 98.4% |
| 8 | 4 | (0.01, 0.05) | 100 | 74.9% | 91.4% | 92.7% | 99.5% | 92.8% | 94.5% | 96.9% | 89.1% | 90.2% | 91.9% | 99.3% | 91.9% | 93.7% | 96.3% | 96.6% |
| 8 | 4 | (0.05, 0.10) | 10 | 90.6% | 91.4% | 93.0% | 99.5% | 93.1% | 94.6% | 97.0% | 92.7% | 90.9% | 92.9% | 99.4% | 93.1% | 94.5% | 96.8% | 97.9% |
| 8 | 4 | (0.05, 0.10) | 100 | 65.6% | 90.8% | 93.1% | 99.8% | 93.1% | 94.8% | 97.3% | 90.2% | 91.1% | 92.3% | 99.6% | 92.3% | 94.0% | 96.9% | 97.0% |
| 8 | 4 | (0.05, 0.15) | 10 | 88.2% | 91.3% | 93.0% | 99.5% | 93.1% | 94.6% | 96.8% | 91.7% | 90.9% | 92.6% | 99.5% | 92.7% | 94.4% | 96.4% | 98.0% |
| 8 | 4 | (0.05, 0.15) | 100 | 55.0% | 90.9% | 92.7% | 99.7% | 92.7% | 94.2% | 96.8% | 88.0% | 90.2% | 92.0% | 99.3% | 92.0% | 93.7% | 96.1% | 95.9% |
| 8 | 8 | (0.01, 0.05) | 10 | 88.3% | 89.2% | 90.8% | - | 90.9% | 92.7% | 95.5% | 89.9% | 89.0% | 90.9% | - | 90.9% | 92.7% | 95.6% | 96.7% |
| 8 | 8 | (0.01, 0.05) | 100 | 62.4% | 90.7% | 92.5% | - | 92.5% | 94.0% | 96.3% | 77.8% | 90.5% | 92.1% | - | 92.1% | 93.7% | 96.0% | 96.0% |
| 8 | 8 | (0.05, 0.10) | 10 | 87.7% | 90.3% | 92.0% | - | 92.2% | 93.5% | 96.2% | 90.3% | 90.2% | 91.8% | - | 91.9% | 93.6% | 95.9% | 96.9% |
| 8 | 8 | (0.05, 0.10) | 100 | 55.7% | 91.9% | 93.7% | - | 93.7% | 94.8% | 96.4% | 81.5% | 91.4% | 92.8% | - | 92.8% | 94.6% | 96.7% | 96.1% |
| 8 | 8 | (0.05, 0.15) | 10 | 82.4% | 90.2% | 92.0% | - | 92.2% | 93.7% | 96.2% | 87.9% | 89.8% | 91.8% | - | 92.0% | 93.7% | 96.1% | 96.8% |
| 8 | 8 | (0.05, 0.15) | 100 | 43.2% | 91.1% | 92.9% | - | 92.9% | 94.4% | 96.2% | 74.9% | 90.7% | 92.6% | - | 92.6% | 94.3% | 96.0% | 95.5% |
| 16 | 4 | (0.01, 0.05) | 10 | 92.1% | 93.8% | 94.4% | 97.4% | 94.5% | 95.4% | 96.3% | 93.5% | 93.7% | 94.4% | 97.4% | 94.5% | 95.2% | 96.4% | 96.9% |
| 16 | 4 | (0.01, 0.05) | 100 | 75.1% | 93.3% | 94.0% | 97.9% | 94.0% | 94.8% | 96.2% | 89.8% | 92.9% | 93.4% | 97.5% | 93.4% | 94.1% | 96.0% | 95.5% |
| 16 | 4 | (0.05, 0.10) | 10 | 90.9% | 93.5% | 94.4% | 97.7% | 94.5% | 95.1% | 96.2% | 93.4% | 93.4% | 94.2% | 97.7% | 94.2% | 94.7% | 96.1% | 96.5% |
| 16 | 4 | (0.05, 0.10) | 100 | 66.4% | 93.2% | 94.1% | 98.0% | 94.1% | 94.8% | 96.2% | 90.9% | 93.1% | 94.1% | 97.6% | 94.1% | 94.4% | 96.2% | 96.3% |
| 16 | 4 | (0.05, 0.15) | 10 | 87.4% | 93.4% | 94.3% | 97.2% | 94.4% | 94.8% | 96.0% | 92.2% | 93.2% | 93.7% | 97.2% | 93.8% | 94.7% | 95.5% | 96.1% |
| 16 | 4 | (0.05, 0.15) | 100 | 54.7% | 94.0% | 94.8% | 97.5% | 94.8% | 95.2% | 96.6% | 89.5% | 93.9% | 94.2% | 97.5% | 94.2% | 95.0% | 96.2% | 96.4% |
| 16 | 8 | (0.01, 0.05) | 10 | 89.5% | 91.9% | 92.6% | 99.2% | 92.7% | 93.5% | 95.0% | 90.7% | 91.8% | 92.5% | 99.2% | 92.5% | 93.6% | 95.1% | 95.7% |
| 16 | 8 | (0.01, 0.05) | 100 | 64.3% | 92.5% | 93.3% | 99.2% | 93.3% | 94.1% | 95.7% | 80.2% | 92.7% | 93.7% | 99.2% | 93.7% | 94.2% | 95.2% | 95.7% |
| 16 | 8 | (0.05, 0.10) | 10 | 87.8% | 92.8% | 93.4% | 99.1% | 93.5% | 94.0% | 95.2% | 90.3% | 92.6% | 93.3% | 99.1% | 93.3% | 94.0% | 95.4% | 95.4% |

| I | S | WPICC (Control, Intervention) | K | Standard | CR0 | CR1 | CR1P | CR1S | CR2 | CR3 | Standard | CR0 | CR1 | CR1P | CR1S | CR2 | CR3 | True model |
|---|---|---|---|---|---|---|---|---|---|---|---|---|---|---|---|---|---|---|
| 16 | 8 | (0.05, 0.10) | 100 | 55.7% | 92.8% | 93.7% | 99.5% | 93.7% | 94.7% | 96.0% | 82.3% | 93.0% | 93.6% | 99.5% | 93.6% | 94.3% | 96.1% | 96.0% |
| 16 | 8 | (0.05, 0.15) | 10 | 82.2% | 92.5% | 93.2% | 99.1% | 93.2% | 93.5% | 94.9% | 87.5% | 92.4% | 93.1% | 99.0% | 93.1% | 93.6% | 94.8% | 95.3% |
| 16 | 8 | (0.05, 0.15) | 100 | 43.0% | 92.5% | 93.4% | 99.4% | 93.4% | 94.1% | 95.8% | 76.2% | 93.1% | 93.4% | 99.2% | 93.4% | 94.2% | 95.5% | 95.8% |
| 32 | 4 | (0.01, 0.05) | 10 | 91.9% | 93.3% | 93.8% | 95.9% | 93.8% | 94.1% | 95.0% | 93.1% | 93.0% | 93.3% | 95.7% | 93.4% | 93.9% | 94.6% | 95.1% |
| 32 | 4 | (0.01, 0.05) | 100 | 73.8% | 93.5% | 93.8% | 96.1% | 93.8% | 94.2% | 95.1% | 89.5% | 93.6% | 94.0% | 96.2% | 94.0% | 94.5% | 95.3% | 94.6% |
| 32 | 4 | (0.05, 0.10) | 10 | 90.8% | 94.2% | 94.6% | 95.9% | 94.6% | 94.9% | 95.4% | 93.5% | 93.9% | 94.3% | 96.1% | 94.3% | 94.6% | 95.2% | 95.4% |
| 32 | 4 | (0.05, 0.10) | 100 | 64.5% | 93.4% | 93.9% | 95.8% | 93.9% | 94.3% | 95.0% | 90.9% | 93.7% | 93.9% | 95.6% | 93.9% | 94.2% | 94.9% | 94.4% |
| 32 | 4 | (0.05, 0.15) | 10 | 87.7% | 93.8% | 94.2% | 95.9% | 94.4% | 94.8% | 95.3% | 92.2% | 93.4% | 93.8% | 95.7% | 93.8% | 94.2% | 95.2% | 95.0% |
| 32 | 4 | (0.05, 0.15) | 100 | 54.1% | 93.5% | 93.9% | 95.6% | 93.9% | 94.0% | 94.8% | 89.3% | 93.5% | 93.9% | 95.5% | 93.9% | 94.2% | 94.9% | 94.6% |
| 32 | 8 | (0.01, 0.05) | 10 | 89.7% | 94.0% | 94.3% | 97.5% | 94.3% | 94.6% | 95.3% | 90.9% | 93.7% | 94.0% | 97.5% | 94.1% | 94.5% | 95.1% | 95.2% |
| 32 | 8 | (0.01, 0.05) | 100 | 66.3% | 93.4% | 93.7% | 97.2% | 93.7% | 94.0% | 94.9% | 80.9% | 93.1% | 93.5% | 97.1% | 93.5% | 94.1% | 94.9% | 95.0% |
| 32 | 8 | (0.05, 0.10) | 10 | 87.5% | 93.4% | 93.7% | 97.5% | 93.7% | 94.2% | 94.7% | 90.5% | 93.3% | 93.8% | 97.4% | 93.8% | 94.1% | 94.6% | 95.1% |
| 32 | 8 | (0.05, 0.10) | 100 | 53.3% | 93.6% | 93.9% | 97.7% | 93.9% | 94.3% | 94.9% | 82.0% | 93.3% | 93.9% | 97.6% | 93.9% | 94.3% | 95.2% | 95.0% |
| 32 | 8 | (0.05, 0.15) | 10 | 82.9% | 93.7% | 94.3% | 97.6% | 94.3% | 94.7% | 95.2% | 87.7% | 93.6% | 94.2% | 97.5% | 94.2% | 94.5% | 95.1% | 95.2% |
| 32 | 8 | (0.05, 0.15) | 100 | 45.8% | 93.5% | 93.8% | 97.2% | 93.8% | 94.1% | 95.0% | 77.4% | 93.3% | 93.6% | 97.4% | 93.6% | 94.1% | 94.8% | 94.5% |

| True model = DTD + Random intervention | | | | EXCH | | | | | | | NE | | | | | | | True model |
|---|---|---|---|---|---|---|---|---|---|---|---|---|---|---|---|---|---|---|
| I | S | WPICC (Control, Intervention) | K | Standard | CR0 | CR1 | CR1P | CR1S | CR2 | CR3 | Standard | CR0 | CR1 | CR1P | CR1S | CR2 | CR3 | |
| 8 | 4 | (0.01, 0.05) | 10 | 92.3% | 90.8% | 92.2% | 99.2% | 92.4% | 93.6% | 96.1% | 93.6% | 90.5% | 92.0% | 99.2% | 92.1% | 93.4% | 95.9% | 97.3% |
| 8 | 4 | (0.01, 0.05) | 100 | 74.4% | 91.1% | 92.6% | 99.5% | 92.6% | 94.2% | 96.7% | 88.6% | 89.7% | 91.8% | 99.4% | 91.8% | 93.6% | 96.3% | 97.1% |
| 8 | 4 | (0.05, 0.10) | 10 | 89.3% | 91.2% | 92.5% | 99.0% | 92.5% | 93.9% | 95.9% | 92.8% | 90.8% | 92.0% | 98.9% | 92.1% | 93.8% | 96.0% | 98.3% |
| 8 | 4 | (0.05, 0.10) | 100 | 63.7% | 92.0% | 93.2% | 99.5% | 93.2% | 94.5% | 97.0% | 90.9% | 91.0% | 92.9% | 99.5% | 93.0% | 94.5% | 96.4% | 96.8% |
| 8 | 4 | (0.05, 0.15) | 10 | 87.1% | 91.3% | 92.6% | 99.1% | 92.7% | 93.8% | 96.0% | 91.2% | 90.6% | 92.1% | 99.0% | 92.2% | 93.8% | 95.7% | 97.7% |
| 8 | 4 | (0.05, 0.15) | 100 | 53.9% | 91.7% | 92.7% | 99.5% | 92.8% | 93.9% | 96.1% | 88.5% | 90.7% | 92.0% | 99.4% | 92.0% | 94.2% | 96.1% | 95.8% |
| 8 | 8 | (0.01, 0.05) | 10 | 88.5% | 90.5% | 92.2% | - | 92.5% | 93.8% | 96.2% | 90.5% | 90.3% | 92.1% | - | 92.3% | 93.6% | 96.1% | 97.9% |
| 8 | 8 | (0.01, 0.05) | 100 | 61.4% | 89.4% | 91.4% | - | 91.5% | 93.2% | 96.1% | 77.7% | 89.4% | 91.1% | - | 91.2% | 92.6% | 95.6% | 96.3% |
| 8 | 8 | (0.05, 0.10) | 10 | 84.0% | 91.0% | 92.3% | - | 92.5% | 94.1% | 96.8% | 89.0% | 90.6% | 92.3% | - | 92.3% | 93.9% | 96.6% | 97.9% |
| 8 | 8 | (0.05, 0.10) | 100 | 48.2% | 91.1% | 92.9% | - | 92.9% | 94.4% | 96.4% | 79.8% | 90.3% | 91.8% | - | 91.8% | 93.8% | 96.4% | 96.8% |
| 8 | 8 | (0.05, 0.15) | 10 | 79.4% | 91.6% | 92.9% | - | 93.1% | 94.1% | 96.9% | 86.9% | 90.5% | 92.1% | - | 92.2% | 93.7% | 96.5% | 97.7% |
| 8 | 8 | (0.05, 0.15) | 100 | 40.4% | 90.8% | 92.7% | - | 92.8% | 94.2% | 96.4% | 75.8% | 89.9% | 91.9% | - | 91.9% | 93.6% | 96.0% | 96.5% |
| 16 | 4 | (0.01, 0.05) | 10 | 91.1% | 91.9% | 93.0% | 97.1% | 93.1% | 93.8% | 95.1% | 92.5% | 92.0% | 92.9% | 97.0% | 92.9% | 93.8% | 94.9% | 96.6% |
| 16 | 4 | (0.01, 0.05) | 100 | 74.3% | 93.5% | 94.4% | 97.4% | 94.4% | 94.8% | 95.9% | 91.3% | 93.4% | 93.9% | 97.2% | 93.9% | 94.7% | 95.7% | 95.9% |
| 16 | 4 | (0.05, 0.10) | 10 | 89.4% | 92.2% | 93.1% | 97.2% | 93.1% | 94.1% | 95.6% | 92.5% | 92.4% | 93.1% | 97.2% | 93.2% | 93.9% | 95.5% | 96.7% |
| 16 | 4 | (0.05, 0.10) | 100 | 63.5% | 93.4% | 94.1% | 97.2% | 94.1% | 94.9% | 96.1% | 91.8% | 93.2% | 93.8% | 97.3% | 93.8% | 94.4% | 95.8% | 95.5% |
| 16 | 4 | (0.05, 0.15) | 10 | 85.8% | 92.4% | 93.3% | 97.4% | 93.4% | 94.1% | 95.5% | 91.2% | 92.1% | 93.0% | 97.1% | 93.1% | 93.9% | 95.4% | 96.1% |
| 16 | 4 | (0.05, 0.15) | 100 | 54.1% | 93.5% | 94.2% | 97.1% | 94.2% | 94.9% | 95.9% | 90.3% | 92.9% | 93.4% | 97.3% | 93.4% | 93.9% | 95.7% | 95.2% |
| 16 | 8 | (0.01, 0.05) | 10 | 89.9% | 92.1% | 92.5% | 99.2% | 92.6% | 93.5% | 94.8% | 91.0% | 92.1% | 92.4% | 99.2% | 92.4% | 93.3% | 94.6% | 96.0% |
| 16 | 8 | (0.01, 0.05) | 100 | 61.2% | 92.2% | 92.8% | 99.6% | 92.8% | 94.0% | 95.4% | 79.6% | 92.1% | 93.0% | 99.6% | 93.0% | 93.7% | 95.3% | 95.9% |
| 16 | 8 | (0.05, 0.10) | 10 | 85.8% | 92.3% | 93.0% | 99.4% | 93.0% | 93.8% | 95.0% | 90.0% | 92.0% | 92.9% | 99.4% | 92.9% | 93.6% | 94.9% | 96.1% |
| 16 | 8 | (0.05, 0.10) | 100 | 46.9% | 93.0% | 93.7% | 99.5% | 93.7% | 94.8% | 96.4% | 82.4% | 92.6% | 93.5% | 99.4% | 93.6% | 94.5% | 96.0% | 95.6% |
| 16 | 8 | (0.05, 0.15) | 10 | 79.6% | 92.1% | 93.2% | 99.2% | 93.3% | 93.9% | 95.4% | 87.0% | 92.1% | 93.0% | 99.3% | 93.1% | 93.5% | 94.9% | 95.8% |
| 16 | 8 | (0.05, 0.15) | 100 | 39.4% | 93.0% | 93.6% | 99.4% | 93.6% | 94.2% | 96.0% | 76.9% | 92.4% | 93.3% | 99.4% | 93.4% | 94.2% | 95.5% | 95.2% |
| 32 | 4 | (0.01, 0.05) | 10 | 92.7% | 94.0% | 94.3% | 96.5% | 94.4% | 94.8% | 95.5% | 93.7% | 94.2% | 94.4% | 96.4% | 94.4% | 94.8% | 95.5% | 95.5% |
| 32 | 4 | (0.01, 0.05) | 100 | 74.0% | 93.4% | 93.9% | 96.0% | 93.9% | 94.4% | 95.2% | 89.6% | 93.5% | 93.9% | 95.4% | 93.9% | 94.3% | 94.8% | 95.3% |
| 32 | 4 | (0.05, 0.10) | 10 | 91.0% | 94.8% | 95.1% | 96.8% | 95.1% | 95.5% | 96.1% | 94.1% | 94.4% | 94.8% | 96.7% | 94.9% | 95.6% | 96.2% | 96.0% |
| 32 | 4 | (0.05, 0.10) | 100 | 62.0% | 94.1% | 94.3% | 95.9% | 94.3% | 94.9% | 95.1% | 92.1% | 93.7% | 94.1% | 95.8% | 94.1% | 94.4% | 95.2% | 94.9% |
| 32 | 4 | (0.05, 0.15) | 10 | 87.8% | 94.5% | 94.8% | 96.8% | 94.9% | 95.4% | 96.4% | 93.1% | 94.5% | 95.2% | 96.5% | 95.3% | 95.6% | 96.1% | 96.1% |
| 32 | 4 | (0.05, 0.15) | 100 | 53.3% | 93.8% | 94.2% | 95.8% | 94.2% | 94.6% | 94.8% | 89.3% | 93.0% | 93.4% | 95.7% | 93.4% | 93.9% | 94.5% | 94.8% |
| 32 | 8 | (0.01, 0.05) | 10 | 90.1% | 94.6% | 94.8% | 97.7% | 94.8% | 95.1% | 95.6% | 91.5% | 94.6% | 94.8% | 97.5% | 94.8% | 95.1% | 95.5% | 95.2% |
| 32 | 8 | (0.01, 0.05) | 100 | 59.6% | 93.2% | 93.5% | 97.5% | 93.5% | 94.0% | 94.9% | 78.2% | 93.0% | 93.3% | 97.4% | 93.3% | 94.0% | 94.8% | 94.8% |
| 32 | 8 | (0.05, 0.10) | 10 | 86.2% | 94.7% | 94.9% | 97.4% | 95.0% | 95.3% | 95.8% | 90.2% | 94.7% | 95.0% | 97.6% | 95.0% | 95.3% | 95.8% | 95.5% |
| 32 | 8 | (0.05, 0.10) | 100 | 47.6% | 94.0% | 94.2% | 97.4% | 94.2% | 94.6% | 95.5% | 81.8% | 93.8% | 94.1% | 97.2% | 94.1% | 94.3% | 95.0% | 94.9% |
| 32 | 8 | (0.05, 0.15) | 10 | 82.3% | 94.6% | 94.8% | 97.2% | 94.8% | 94.9% | 95.4% | 88.4% | 94.4% | 94.5% | 97.2% | 94.5% | 94.9% | 95.4% | 95.2% |
| 32 | 8 | (0.05, 0.15) | 100 | 38.7% | 93.4% | 93.8% | 97.3% | 93.8% | 94.2% | 94.9% | 76.2% | 93.0% | 93.5% | 96.9% | 93.5% | 94.1% | 94.8% | 94.5% |

# Table S3 Percentage error of model-based SE to empirical SE (%)

| | | True model = NE + Random intervention | | EXCH | | | | | | | NE | | | | | | | True model |
|---|---|---|---|---|---|---|---|---|---|---|---|---|---|---|---|---|---|---|
| I | S | WPICC (Control, Intervention) | K | Standard | CR0 | CR1 | CR1P | CR1S | CR2 | CR3 | Standard | CR0 | CR1 | CR1P | CR1S | CR2 | CR3 | |
| 8 | 4 | (0.01, 0.05) | 10 | -10.9 | -20.6 | -15.1 | 58.8 | -14.6 | -7.7 | 7.3 | -6.1 | -21.3 | -15.9 | 57.3 | -15.4 | -8.5 | 6.5 | 0.9 |
| 8 | 4 | (0.01, 0.05) | 100 | -41.6 | -20.6 | -15.1 | 58.9 | -15.0 | -8.4 | 5.7 | -18.6 | -22.6 | -17.3 | 54.8 | -17.2 | -10.5 | 3.3 | -2.7 |
| 8 | 4 | (0.05, 0.10) | 10 | -14.9 | -19.9 | -14.4 | 60.1 | -13.9 | -7.3 | 7.5 | -7.7 | -21.2 | -15.8 | 57.5 | -15.3 | -8.6 | 6.1 | -0.1 |
| 8 | 4 | (0.05, 0.10) | 100 | -51.4 | -18.9 | -13.3 | 62.2 | -13.2 | -6.5 | 7.9 | -15.0 | -20.9 | -15.4 | 58.2 | -15.4 | -8.6 | 5.5 | -2.4 |
| 8 | 4 | (0.05, 0.15) | 10 | -21.1 | -19.4 | -13.9 | 61.1 | -13.3 | -6.8 | 7.9 | -11.3 | -21.2 | -15.7 | 57.6 | -15.2 | -8.6 | 5.8 | 0.2 |
| 8 | 4 | (0.05, 0.15) | 100 | -61.3 | -19.6 | -14.0 | 60.8 | -14.0 | -7.4 | 6.8 | -19.9 | -22.1 | -16.7 | 55.8 | -16.6 | -10.0 | 3.7 | -2.7 |
| 8 | 8 | (0.01, 0.05) | 10 | -19.5 | -23.8 | -18.6 | - | -18.1 | -11.7 | 2.3 | -15.6 | -24.3 | -19.1 | - | -18.6 | -12.2 | 1.8 | -4.2 |
| 8 | 8 | (0.01, 0.05) | 100 | -55.0 | -23.5 | -18.2 | - | -18.1 | -11.9 | 1.5 | -37.6 | -24.2 | -18.9 | - | -18.9 | -12.6 | 0.5 | -5.0 |
| 8 | 8 | (0.05, 0.10) | 10 | -22.7 | -21.8 | -16.4 | - | -15.8 | -9.6 | 4.4 | -16.3 | -22.6 | -17.2 | - | -16.7 | -10.4 | 3.5 | -3.7 |
| 8 | 8 | (0.05, 0.10) | 100 | -62.1 | -22.0 | -16.6 | - | -16.5 | -10.1 | 3.6 | -34.0 | -22.5 | -17.1 | - | -17.1 | -10.6 | 2.9 | -4.8 |
| 8 | 8 | (0.05, 0.15) | 10 | -32.1 | -21.6 | -16.2 | - | -15.6 | -9.6 | 4.3 | -23.6 | -22.6 | -17.3 | - | -16.7 | -10.6 | 3.1 | -3.6 |
| 8 | 8 | (0.05, 0.15) | 100 | -71.5 | -22.9 | -17.6 | - | -17.6 | -11.3 | 2.1 | -42.0 | -23.6 | -18.4 | - | -18.3 | -12.0 | 1.2 | -5.2 |
| 16 | 4 | (0.01, 0.05) | 10 | -8.8 | -9.2 | -6.2 | 14.9 | -5.9 | -2.6 | 4.4 | -4.3 | -9.4 | -6.4 | 14.7 | -6.1 | -2.8 | 4.3 | -0.4 |
| 16 | 4 | (0.01, 0.05) | 100 | -41.6 | -9.9 | -6.9 | 14.0 | -6.9 | -3.7 | 2.9 | -16.8 | -10.7 | -7.8 | 12.9 | -7.8 | -4.5 | 2.0 | -2.1 |
| 16 | 4 | (0.05, 0.10) | 10 | -14.8 | -10.1 | -7.2 | 13.7 | -6.9 | -3.8 | 3.0 | -7.2 | -10.6 | -7.6 | 13.1 | -7.3 | -4.2 | 2.6 | -2.3 |
| 16 | 4 | (0.05, 0.10) | 100 | -51.2 | -8.2 | -5.2 | 16.1 | -5.2 | -2.0 | 4.8 | -12.6 | -8.9 | -5.9 | 15.2 | -5.9 | -2.6 | 4.0 | -2.1 |
| 16 | 4 | (0.05, 0.15) | 10 | -22.8 | -12.0 | -9.1 | 11.3 | -8.8 | -5.8 | 0.8 | -12.8 | -13.0 | -10.2 | 10.0 | -9.9 | -6.8 | -0.3 | -3.7 |
| 16 | 4 | (0.05, 0.15) | 100 | -61.0 | -8.3 | -5.3 | 16.0 | -5.3 | -2.1 | 4.6 | -17.4 | -9.7 | -6.8 | 14.2 | -6.7 | -3.4 | 3.1 | -0.8 |
| 16 | 8 | (0.01, 0.05) | 10 | -17.7 | -11.9 | -9.0 | 43.9 | -8.7 | -5.7 | 1.0 | -14.1 | -12.1 | -9.2 | 43.5 | -8.9 | -5.8 | 0.8 | -2.0 |
| 16 | 8 | (0.01, 0.05) | 100 | -53.5 | -10.8 | -7.9 | 45.7 | -7.8 | -4.8 | 1.7 | -34.9 | -11.0 | -8.0 | 45.4 | -8.0 | -4.9 | 1.5 | -0.8 |
| 16 | 8 | (0.05, 0.10) | 10 | -21.8 | -11.0 | -8.1 | 45.4 | -7.8 | -4.8 | 1.8 | -15.1 | -11.3 | -8.4 | 44.8 | -8.1 | -5.1 | 1.5 | -2.0 |
| 16 | 8 | (0.05, 0.10) | 100 | -60.9 | -9.9 | -7.0 | 47.1 | -6.9 | -3.8 | 2.7 | -31.3 | -9.9 | -6.9 | 47.1 | -6.9 | -3.7 | 2.8 | -1.0 |

| I | S | WPICC (Control, Intervention) | K | Standard | CR0 | CR1 | CR1P | CR1S | CR2 | CR3 | Standard | CR0 | CR1 | CR1P | CR1S | CR2 | CR3 | True model |
|---|---|---|---|---|---|---|---|---|---|---|---|---|---|---|---|---|---|---|
| 16 | 8 | (0.05, 0.15) | 10 | -31.9 | -11.0 | -8.1 | 45.3 | -7.8 | -5.0 | 1.6 | -22.8 | -11.4 | -8.5 | 44.6 | -8.3 | -5.3 | 1.1 | -1.7 |
| 16 | 8 | (0.05, 0.15) | 100 | -70.6 | -10.7 | -7.7 | 45.9 | -7.7 | -4.6 | 1.8 | -39.5 | -10.7 | -7.8 | 45.8 | -7.8 | -4.7 | 1.7 | -1.7 |
| 32 | 4 | (0.01, 0.05) | 10 | -11.1 | -6.7 | -5.2 | 3.5 | -5.1 | -3.6 | -0.3 | -7.3 | -7.1 | -5.6 | 3.1 | -5.4 | -3.9 | -0.6 | -1.9 |
| 32 | 4 | (0.01, 0.05) | 100 | -42.3 | -6.1 | -4.6 | 4.2 | -4.6 | -3.0 | 0.1 | -16.7 | -6.2 | -4.7 | 4.1 | -4.7 | -3.1 | 0.1 | -1.9 |
| 32 | 4 | (0.05, 0.10) | 10 | -14.1 | -5.5 | -4.0 | 4.9 | -3.8 | -2.3 | 0.9 | -6.7 | -5.9 | -4.4 | 4.4 | -4.2 | -2.7 | 0.6 | -1.2 |
| 32 | 4 | (0.05, 0.10) | 100 | -52.7 | -6.2 | -4.7 | 4.0 | -4.7 | -3.2 | 0.0 | -14.4 | -6.3 | -4.8 | 3.9 | -4.8 | -3.2 | -0.1 | -2.9 |
| 32 | 4 | (0.05, 0.15) | 10 | -21.4 | -6.0 | -4.4 | 4.3 | -4.3 | -2.8 | 0.4 | -10.6 | -6.5 | -5.0 | 3.8 | -4.8 | -3.3 | -0.1 | -1.6 |
| 32 | 4 | (0.05, 0.15) | 100 | -62.6 | -7.2 | -5.7 | 3.0 | -5.7 | -4.2 | -1.1 | -19.4 | -7.3 | -5.8 | 2.8 | -5.8 | -4.3 | -1.2 | -3.8 |
| 32 | 8 | (0.01, 0.05) | 10 | -15.3 | -4.8 | -3.3 | 14.8 | -3.1 | -1.6 | 1.7 | -12.2 | -5.2 | -3.7 | 14.3 | -3.5 | -2.0 | 1.3 | -0.2 |
| 32 | 8 | (0.01, 0.05) | 100 | -53.2 | -4.8 | -3.3 | 14.8 | -3.3 | -1.8 | 1.4 | -34.2 | -5.3 | -3.8 | 14.2 | -3.8 | -2.3 | 0.9 | -1.5 |
| 32 | 8 | (0.05, 0.10) | 10 | -20.9 | -5.6 | -4.1 | 13.8 | -4.0 | -2.5 | 0.7 | -14.4 | -6.0 | -4.5 | 13.4 | -4.3 | -2.9 | 0.3 | -1.2 |
| 32 | 8 | (0.05, 0.10) | 100 | -62.2 | -7.2 | -5.8 | 11.9 | -5.7 | -4.2 | -1.2 | -33.0 | -7.2 | -5.7 | 11.9 | -5.7 | -4.2 | -1.1 | -3.6 |
| 32 | 8 | (0.05, 0.15) | 10 | -29.9 | -4.2 | -2.6 | 15.6 | -2.5 | -1.0 | 2.2 | -20.9 | -4.6 | -3.1 | 15.0 | -2.9 | -1.5 | 1.7 | 0.0 |
| 32 | 8 | (0.05, 0.15) | 100 | -70.6 | -5.1 | -3.5 | 14.5 | -3.5 | -2.0 | 1.1 | -39.1 | -5.6 | -4.0 | 13.9 | -4.0 | -2.5 | 0.6 | 0.0 |

| True model = DTD + Random intervention | | | | EXCH | | | | | | | NE | | | | | | | True model |
|---|---|---|---|---|---|---|---|---|---|---|---|---|---|---|---|---|---|---|
| I | S | WPICC (Control, Intervention) | K | Standard | CR0 | CR1 | CR1P | CR1S | CR2 | CR3 | Standard | CR0 | CR1 | CR1P | CR1S | CR2 | CR3 | |
| 8 | 4 | (0.01, 0.05) | 10 | -12.2 | -21.5 | -16.1 | 56.9 | -15.6 | -8.8 | 6.1 | -6.9 | -22.0 | -16.6 | 56.0 | -16.1 | -9.1 | 5.8 | 2.2 |
| 8 | 4 | (0.01, 0.05) | 100 | -43.1 | -20.6 | -15.1 | 58.9 | -15.0 | -8.4 | 5.8 | -18.4 | -22.3 | -17.0 | 55.3 | -16.9 | -10.0 | 3.9 | -1.8 |
| 8 | 4 | (0.05, 0.10) | 10 | -17.3 | -20.8 | -15.3 | 58.5 | -14.8 | -8.1 | 6.6 | -8.4 | -21.4 | -16.0 | 57.2 | -15.5 | -8.7 | 6.0 | 2.0 |
| 8 | 4 | (0.05, 0.10) | 100 | -54.2 | -19.3 | -13.8 | 61.3 | -13.7 | -7.0 | 7.4 | -14.1 | -21.4 | -16.0 | 57.2 | -15.9 | -8.9 | 5.1 | -1.1 |
| 8 | 4 | (0.05, 0.15) | 10 | -23.7 | -21.2 | -15.7 | 57.7 | -15.2 | -8.7 | 5.7 | -12.6 | -22.2 | -16.9 | 55.5 | -16.3 | -9.7 | 4.7 | 0.4 |
| 8 | 4 | (0.05, 0.15) | 100 | -62.7 | -19.9 | -14.4 | 60.2 | -14.3 | -7.7 | 6.5 | -18.7 | -21.9 | -16.5 | 56.2 | -16.5 | -9.6 | 4.2 | -1.6 |
| 8 | 8 | (0.01, 0.05) | 10 | -20.5 | -22.9 | -17.6 | - | -17.0 | -10.6 | 3.8 | -16.0 | -23.3 | -18.0 | - | -17.5 | -10.9 | 3.5 | -0.7 |
| 8 | 8 | (0.01, 0.05) | 100 | -56.7 | -23.0 | -17.7 | - | -17.6 | -11.3 | 2.3 | -37.5 | -24.4 | -19.2 | - | -19.1 | -12.8 | 0.5 | -3.4 |
| 8 | 8 | (0.05, 0.10) | 10 | -28.2 | -20.7 | -15.2 | - | -14.7 | -8.3 | 6.2 | -18.9 | -21.3 | -15.9 | - | -15.4 | -8.9 | 5.7 | -0.1 |

| | | | | | | | | | | | | | | | | | |
|---|---|---|---|---|---|---|---|---|---|---|---|---|---|---|---|---|---|
| 8 | 8 | (0.05, 0.10) | 100 | -67.5 | -21.2 | -15.8 | - | -15.7 | -9.1 | 4.9 | -34.4 | -22.9 | -17.6 | - | -17.6 | -11.0 | 2.7 | -3.1 |
| 8 | 8 | (0.05, 0.15) | 10 | -36.0 | -20.9 | -15.4 | - | -14.9 | -8.7 | 5.5 | -25.1 | -21.8 | -16.4 | - | -15.9 | -9.6 | 4.5 | -1.0 |
| 8 | 8 | (0.05, 0.15) | 100 | -73.6 | -21.8 | -16.4 | - | -16.4 | -9.9 | 3.9 | -40.4 | -23.6 | -18.3 | - | -18.2 | -11.9 | 1.6 | -3.5 |
| 16 | 4 | (0.01, 0.05) | 10 | -13.4 | -12.9 | -10.0 | 10.2 | -9.7 | -6.6 | 0.2 | -9.0 | -13.3 | -10.4 | 9.7 | -10.1 | -6.9 | -0.2 | 1.6 |
| 16 | 4 | (0.01, 0.05) | 100 | -41.9 | -8.4 | -5.3 | 15.9 | -5.3 | -2.1 | 4.7 | -14.4 | -8.5 | -5.5 | 15.7 | -5.5 | -2.1 | 4.6 | 0.2 |
| 16 | 4 | (0.05, 0.10) | 10 | -17.6 | -11.8 | -8.9 | 11.6 | -8.6 | -5.5 | 1.1 | -9.2 | -12.6 | -9.8 | 10.5 | -9.5 | -6.3 | 0.3 | -0.1 |
| 16 | 4 | (0.05, 0.10) | 100 | -54.1 | -9.2 | -6.2 | 14.9 | -6.2 | -2.9 | 3.7 | -11.1 | -9.6 | -6.6 | 14.4 | -6.6 | -3.2 | 3.5 | -0.8 |
| 16 | 4 | (0.05, 0.15) | 10 | -23.7 | -11.6 | -8.7 | 11.8 | -8.5 | -5.5 | 1.2 | -12.4 | -12.6 | -9.8 | 10.5 | -9.5 | -6.4 | 0.2 | -0.5 |
| 16 | 4 | (0.05, 0.15) | 100 | -62.7 | -9.5 | -6.5 | 14.5 | -6.5 | -3.4 | 3.2 | -16.0 | -9.9 | -7.0 | 14.0 | -6.9 | -3.6 | 2.9 | -1.1 |
| 16 | 8 | (0.01, 0.05) | 10 | -18.7 | -11.6 | -8.8 | 44.3 | -8.5 | -5.4 | 1.3 | -14.6 | -12.0 | -9.1 | 43.7 | -8.8 | -5.7 | 1.0 | 0.3 |
| 16 | 8 | (0.01, 0.05) | 100 | -56.0 | -10.5 | -7.6 | 46.2 | -7.5 | -4.4 | 2.1 | -35.0 | -11.1 | -8.2 | 45.1 | -8.2 | -5.0 | 1.4 | -0.8 |
| 16 | 8 | (0.05, 0.10) | 10 | -26.6 | -10.1 | -7.2 | 46.8 | -6.9 | -3.9 | 2.8 | -17.2 | -10.7 | -7.8 | 45.8 | -7.5 | -4.4 | 2.3 | 0.0 |
| 16 | 8 | (0.05, 0.10) | 100 | -67.1 | -9.4 | -6.4 | 48.0 | -6.4 | -3.2 | 3.5 | -31.9 | -10.1 | -7.2 | 46.8 | -7.1 | -3.9 | 2.7 | -1.1 |
| 16 | 8 | (0.05, 0.15) | 10 | -34.9 | -10.3 | -7.3 | 46.5 | -7.1 | -4.1 | 2.5 | -23.5 | -11.0 | -8.0 | 45.4 | -7.8 | -4.8 | 1.8 | -0.1 |
| 16 | 8 | (0.05, 0.15) | 100 | -73.5 | -10.1 | -7.2 | 46.7 | -7.2 | -4.0 | 2.5 | -38.3 | -11.0 | -8.0 | 45.4 | -8.0 | -4.8 | 1.6 | -1.4 |
| 32 | 4 | (0.01, 0.05) | 10 | -11.0 | -5.8 | -4.3 | 4.5 | -4.2 | -2.6 | 0.7 | -6.7 | -5.9 | -4.4 | 4.4 | -4.3 | -2.7 | 0.7 | -0.8 |
| 32 | 4 | (0.01, 0.05) | 100 | -42.8 | -6.2 | -4.7 | 4.0 | -4.7 | -3.2 | 0.0 | -16.6 | -7.3 | -5.8 | 2.8 | -5.8 | -4.2 | -1.1 | -2.2 |
| 32 | 4 | (0.05, 0.10) | 10 | -14.9 | -4.1 | -2.6 | 6.3 | -2.5 | -0.9 | 2.4 | -5.8 | -4.4 | -2.8 | 6.1 | -2.7 | -1.1 | 2.2 | 1.1 |
| 32 | 4 | (0.05, 0.10) | 100 | -54.4 | -5.7 | -4.2 | 4.7 | -4.1 | -2.6 | 0.6 | -11.9 | -6.6 | -5.1 | 3.6 | -5.1 | -3.4 | -0.3 | -1.4 |
| 32 | 4 | (0.05, 0.15) | 10 | -21.9 | -4.7 | -3.1 | 5.8 | -3.0 | -1.5 | 1.8 | -9.5 | -4.9 | -3.4 | 5.5 | -3.2 | -1.7 | 1.6 | 0.1 |
| 32 | 4 | (0.05, 0.15) | 100 | -63.0 | -6.1 | -4.6 | 4.1 | -4.6 | -3.1 | 0.0 | -17.0 | -7.1 | -5.6 | 3.1 | -5.6 | -4.0 | -0.9 | -2.0 |
| 32 | 8 | (0.01, 0.05) | 10 | -16.0 | -3.9 | -2.4 | 15.9 | -2.3 | -0.7 | 2.6 | -12.2 | -4.2 | -2.6 | 15.6 | -2.5 | -0.9 | 2.4 | 1.5 |
| 32 | 8 | (0.01, 0.05) | 100 | -57.3 | -7.9 | -6.4 | 11.1 | -6.4 | -4.9 | -1.8 | -36.3 | -8.3 | -6.8 | 10.6 | -6.8 | -5.3 | -2.3 | -3.5 |
| 32 | 8 | (0.05, 0.10) | 10 | -25.0 | -3.5 | -2.0 | 16.3 | -1.8 | -0.4 | 2.9 | -15.4 | -3.9 | -2.3 | 15.9 | -2.2 | -0.7 | 2.6 | 0.4 |
| 32 | 8 | (0.05, 0.10) | 100 | -67.8 | -6.7 | -5.2 | 12.5 | -5.2 | -3.7 | -0.6 | -32.9 | -7.4 | -5.9 | 11.7 | -5.9 | -4.4 | -1.3 | -2.2 |
| 32 | 8 | (0.05, 0.15) | 10 | -33.1 | -3.2 | -1.6 | 16.8 | -1.4 | 0.0 | 3.3 | -21.3 | -3.4 | -1.9 | 16.5 | -1.7 | -0.2 | 3.0 | 1.1 |
| 32 | 8 | (0.05, 0.15) | 100 | -74.1 | -7.3 | -5.8 | 11.8 | -5.8 | -4.3 | -1.2 | -39.2 | -7.9 | -6.4 | 11.1 | -6.4 | -4.9 | -1.9 | -3.1 |

Table S4 Coverage probability of 95% confidence interval (Degree of Freedom: Satterthwaite)

| | | True model = NE + Random intervention | | EXCH | | | | | | | NE | | | | | | | True model* |
|---|---|---|---|---|---|---|---|---|---|---|---|---|---|---|---|---|---|---|
| I | S | WPICC (Control, Intervention) | K | Standard | CR0 | CR1 | CR1P | CR1S | CR2 | CR3 | Standard | CR0 | CR1 | CR1P | CR1S | CR2 | CR3 | |
| 8 | 4 | (0.01, 0.05) | 10 | 92.0% | 85.4% | 88.1% | 98.7% | 88.4% | 90.5% | 94.1% | 93.9% | 85.7% | 88.5% | 98.7% | 88.8% | 90.8% | 94.1% | 94.5% |
| 8 | 4 | (0.01, 0.05) | 100 | 74.9% | 84.0% | 86.4% | 98.3% | 86.4% | 89.0% | 92.7% | 89.1% | 84.3% | 86.9% | 98.3% | 86.9% | 89.4% | 92.7% | 92.7% |
| 8 | 4 | (0.05, 0.10) | 10 | 90.6% | 86.1% | 87.5% | 98.6% | 87.7% | 89.7% | 93.4% | 92.7% | 85.8% | 87.4% | 98.6% | 87.7% | 90.0% | 93.8% | 93.2% |
| 8 | 4 | (0.05, 0.10) | 100 | 65.6% | 84.6% | 86.6% | 98.7% | 86.6% | 88.8% | 93.1% | 90.2% | 85.2% | 86.7% | 98.6% | 86.8% | 89.6% | 93.0% | 92.2% |
| 8 | 4 | (0.05, 0.15) | 10 | 88.2% | 85.2% | 87.5% | 98.6% | 87.7% | 89.7% | 93.1% | 91.7% | 85.7% | 87.7% | 98.5% | 87.9% | 89.8% | 93.5% | 93.3% |
| 8 | 4 | (0.05, 0.15) | 100 | 55.0% | 84.2% | 86.5% | 98.6% | 86.5% | 88.6% | 92.6% | 88.0% | 84.7% | 86.9% | 98.7% | 86.9% | 89.4% | 92.9% | 91.6% |
| 8 | 8 | (0.01, 0.05) | 10 | 88.3% | 83.3% | 85.5% | | 85.6% | 87.3% | 91.3% | 89.9% | 83.1% | 85.6% | | 85.8% | 87.4% | 91.7% | 92.5% |
| 8 | 8 | (0.01, 0.05) | 100 | 62.4% | 84.1% | 86.5% | | 86.5% | 88.8% | 92.3% | 77.8% | 84.2% | 86.6% | | 86.6% | 88.6% | 92.4% | 91.6% |
| 8 | 8 | (0.05, 0.10) | 10 | 87.7% | 83.8% | 86.2% | | 86.6% | 88.7% | 92.3% | 90.3% | 84.0% | 86.4% | | 86.8% | 88.6% | 92.5% | 92.6% |
| 8 | 8 | (0.05, 0.10) | 100 | 55.7% | 85.5% | 87.6% | | 87.6% | 89.9% | 93.7% | 81.5% | 85.6% | 87.6% | | 87.6% | 90.2% | 93.2% | 92.4% |
| 8 | 8 | (0.05, 0.15) | 10 | 82.4% | 83.4% | 85.9% | | 86.2% | 88.1% | 91.8% | 87.9% | 83.7% | 86.3% | | 86.5% | 88.3% | 92.2% | 92.4% |
| 8 | 8 | (0.05, 0.15) | 100 | 43.2% | 84.9% | 87.0% | | 87.0% | 88.9% | 92.9% | 74.9% | 84.7% | 87.0% | | 87.0% | 89.3% | 93.1% | 92.2% |
| 16 | 4 | (0.01, 0.05) | 10 | 92.1% | 91.5% | 92.5% | 96.5% | 92.5% | 93.5% | 95.0% | 93.5% | 91.6% | 92.5% | 96.6% | 92.6% | 93.6% | 95.1% | 95.1% |
| 16 | 4 | (0.01, 0.05) | 100 | 75.1% | 90.5% | 91.3% | 96.5% | 91.3% | 92.6% | 94.2% | 89.8% | 90.8% | 92.0% | 96.2% | 92.0% | 92.7% | 94.1% | 94.0% |
| 16 | 4 | (0.05, 0.10) | 10 | 90.9% | 90.7% | 91.9% | 96.3% | 92.1% | 93.1% | 94.5% | 93.4% | 90.9% | 92.1% | 96.6% | 92.3% | 93.3% | 94.5% | 94.3% |
| 16 | 4 | (0.05, 0.10) | 100 | 66.4% | 90.8% | 91.5% | 96.4% | 91.5% | 92.4% | 94.4% | 90.9% | 91.2% | 92.0% | 96.7% | 92.0% | 93.1% | 94.4% | 93.8% |
| 16 | 4 | (0.05, 0.15) | 10 | 87.4% | 91.1% | 92.2% | 96.1% | 92.3% | 92.8% | 94.6% | 92.2% | 91.4% | 91.9% | 96.2% | 92.0% | 93.2% | 94.7% | 94.1% |
| 16 | 4 | (0.05, 0.15) | 100 | 54.7% | 91.5% | 92.4% | 96.7% | 92.4% | 93.6% | 95.0% | 89.5% | 91.2% | 92.4% | 96.7% | 92.4% | 93.8% | 94.7% | 94.4% |
| 16 | 8 | (0.01, 0.05) | 10 | 89.5% | 90.0% | 90.7% | 98.5% | 90.9% | 91.5% | 93.2% | 90.7% | 90.0% | 90.7% | 98.4% | 90.8% | 91.6% | 93.4% | 94.0% |
| 16 | 8 | (0.01, 0.05) | 100 | 64.3% | 90.0% | 90.9% | 98.7% | 90.9% | 91.9% | 93.5% | 80.2% | 90.5% | 91.3% | 98.7% | 91.3% | 92.3% | 94.0% | 93.7% |
| 16 | 8 | (0.05, 0.10) | 10 | 87.8% | 90.0% | 91.2% | 98.7% | 91.2% | 92.2% | 93.7% | 90.3% | 90.3% | 91.3% | 98.6% | 91.4% | 92.5% | 93.8% | 93.7% |
| 16 | 8 | (0.05, 0.10) | 100 | 55.7% | 90.8% | 91.1% | 99.0% | 91.1% | 92.0% | 94.1% | 82.3% | 91.2% | 91.6% | 98.9% | 91.6% | 92.7% | 94.0% | 94.1% |

| I | S | WPICC (Control, Intervention) | K | Standard | CR0 | CR1 | CR1P | CR1S | CR2 | CR3 | Standard | CR0 | CR1 | CR1P | CR1S | CR2 | CR3 | True model* |
|---|---|---|---|---|---|---|---|---|---|---|---|---|---|---|---|---|---|---|
| 16 | 8 | (0.05, 0.15) | 10 | 82.2% | 90.3% | 91.4% | 98.3% | 91.4% | 92.2% | 93.2% | 87.5% | 90.5% | 91.2% | 98.4% | 91.3% | 92.1% | 93.3% | 93.2% |
| 16 | 8 | (0.05, 0.15) | 100 | 43.0% | 90.3% | 91.1% | 98.8% | 91.1% | 92.0% | 93.6% | 76.2% | 90.5% | 91.8% | 98.7% | 91.8% | 92.6% | 93.9% | 94.0% |
| 32 | 4 | (0.01, 0.05) | 10 | 91.9% | 92.2% | 92.8% | 95.0% | 92.9% | 93.3% | 94.1% | 93.1% | 92.0% | 92.3% | 94.8% | 92.5% | 93.0% | 93.9% | 93.9% |
| 32 | 4 | (0.01, 0.05) | 100 | 73.8% | 92.6% | 93.0% | 95.1% | 93.0% | 93.3% | 94.1% | 89.5% | 92.7% | 93.1% | 95.5% | 93.1% | 93.6% | 94.4% | 94.0% |
| 32 | 4 | (0.05, 0.10) | 10 | 90.8% | 93.1% | 93.6% | 95.4% | 93.6% | 94.2% | 94.8% | 93.5% | 92.9% | 93.5% | 95.4% | 93.6% | 93.9% | 94.6% | 94.3% |
| 32 | 4 | (0.05, 0.10) | 100 | 64.5% | 92.5% | 92.9% | 95.0% | 92.9% | 93.3% | 94.1% | 90.9% | 92.9% | 93.1% | 95.0% | 93.1% | 93.7% | 94.2% | 93.6% |
| 32 | 4 | (0.05, 0.15) | 10 | 87.7% | 92.8% | 93.2% | 95.2% | 93.3% | 93.5% | 94.7% | 92.2% | 92.9% | 93.2% | 95.2% | 93.3% | 93.4% | 94.2% | 94.1% |
| 32 | 4 | (0.05, 0.15) | 100 | 54.1% | 92.5% | 92.7% | 94.9% | 92.7% | 93.4% | 94.0% | 89.3% | 92.8% | 93.1% | 95.0% | 93.1% | 93.6% | 94.2% | 93.7% |
| 32 | 8 | (0.01, 0.05) | 10 | 89.7% | 92.5% | 93.1% | 96.9% | 93.1% | 93.5% | 94.6% | 90.9% | 92.4% | 93.1% | 96.9% | 93.2% | 93.6% | 94.4% | 94.8% |
| 32 | 8 | (0.01, 0.05) | 100 | 66.3% | 92.2% | 92.6% | 96.4% | 92.6% | 93.2% | 93.8% | 80.9% | 92.2% | 92.8% | 96.4% | 92.8% | 93.0% | 93.8% | 93.4% |
| 32 | 8 | (0.05, 0.10) | 10 | 87.5% | 92.3% | 92.6% | 96.7% | 92.7% | 93.2% | 93.9% | 90.5% | 92.5% | 92.8% | 96.9% | 92.8% | 93.3% | 94.1% | 94.2% |
| 32 | 8 | (0.05, 0.10) | 100 | 53.3% | 92.2% | 92.7% | 96.8% | 92.7% | 93.4% | 94.0% | 82.0% | 92.7% | 92.9% | 97.2% | 92.9% | 93.2% | 94.1% | 93.9% |
| 32 | 8 | (0.05, 0.15) | 10 | 82.9% | 92.5% | 92.7% | 96.8% | 92.8% | 93.3% | 94.5% | 87.7% | 92.5% | 92.9% | 96.8% | 92.9% | 93.6% | 94.4% | 94.2% |
| 32 | 8 | (0.05, 0.15) | 100 | 45.8% | 92.2% | 92.8% | 96.9% | 92.8% | 93.1% | 93.9% | 77.4% | 92.2% | 92.6% | 96.9% | 92.7% | 93.2% | 93.9% | 93.7% |

| True model = DTD + Random intervention | | | | EXCH | | | | | | | NE | | | | | | | True model* |
|---|---|---|---|---|---|---|---|---|---|---|---|---|---|---|---|---|---|---|
| I | S | WPICC (Control, Intervention) | K | Standard | CR0 | CR1 | CR1P | CR1S | CR2 | CR3 | Standard | CR0 | CR1 | CR1P | CR1S | CR2 | CR3 | |
| 8 | 4 | (0.01, 0.05) | 10 | 92.3% | 85.2% | 87.0% | 98.2% | 87.2% | 89.5% | 92.4% | 93.6% | 85.2% | 87.3% | 98.3% | 87.5% | 89.6% | 92.4% | 94.2% |
| 8 | 4 | (0.01, 0.05) | 100 | 74.4% | 84.3% | 86.6% | 98.5% | 86.6% | 89.2% | 92.6% | 88.6% | 85.1% | 86.9% | 98.5% | 87.0% | 88.7% | 92.6% | 92.9% |
| 8 | 4 | (0.05, 0.10) | 10 | 89.3% | 85.2% | 87.9% | 97.6% | 88.0% | 89.8% | 92.8% | 92.8% | 85.5% | 87.7% | 97.9% | 87.8% | 89.9% | 92.8% | 94.2% |
| 8 | 4 | (0.05, 0.10) | 100 | 63.7% | 85.6% | 87.9% | 98.8% | 87.9% | 90.0% | 93.3% | 90.9% | 86.0% | 87.6% | 98.6% | 87.6% | 90.1% | 93.9% | 92.6% |
| 8 | 4 | (0.05, 0.15) | 10 | 87.1% | 85.6% | 87.8% | 97.7% | 88.1% | 90.1% | 92.7% | 91.2% | 85.5% | 87.7% | 98.0% | 88.0% | 90.0% | 93.0% | 93.6% |
| 8 | 4 | (0.05, 0.15) | 100 | 53.9% | 86.1% | 87.6% | 98.5% | 87.7% | 90.0% | 92.7% | 88.5% | 85.8% | 87.6% | 98.2% | 87.6% | 89.6% | 92.9% | 91.7% |
| 8 | 8 | (0.01, 0.05) | 10 | 88.5% | 83.9% | 86.6% | | 86.7% | 89.1% | 92.8% | 90.5% | 84.2% | 86.5% | | 86.7% | 89.0% | 92.7% | 94.2% |
| 8 | 8 | (0.01, 0.05) | 100 | 61.4% | 82.5% | 85.1% | | 85.1% | 87.0% | 91.4% | 77.7% | 82.7% | 85.2% | | 85.2% | 87.5% | 91.3% | 92.0% |
| 8 | 8 | (0.05, 0.10) | 10 | 84.0% | 84.8% | 87.1% | | 87.3% | 89.1% | 92.5% | 89.0% | 85.1% | 87.0% | | 87.2% | 89.2% | 92.8% | 93.8% |
| 8 | 8 | (0.05, 0.10) | 100 | 48.2% | 84.4% | 86.4% | | 86.4% | 88.5% | 92.6% | 79.8% | 84.2% | 86.3% | | 86.4% | 88.1% | 92.5% | 92.0% |
| 8 | 8 | (0.05, 0.15) | 10 | 79.4% | 84.5% | 86.7% | | 86.9% | 89.1% | 92.8% | 86.9% | 84.6% | 87.0% | | 87.2% | 88.9% | 92.5% | 93.7% |
| 8 | 8 | (0.05, 0.15) | 100 | 40.4% | 83.6% | 86.1% | | 86.1% | 88.7% | 92.7% | 75.8% | 83.0% | 85.5% | | 85.5% | 88.4% | 92.2% | 91.2% |
| 16 | 4 | (0.01, 0.05) | 10 | 91.1% | 89.6% | 90.5% | 95.4% | 90.5% | 91.6% | 93.6% | 92.5% | 89.3% | 90.5% | 95.6% | 90.6% | 91.9% | 93.8% | 95.3% |
| 16 | 4 | (0.01, 0.05) | 100 | 74.3% | 91.7% | 92.4% | 96.1% | 92.4% | 93.0% | 94.6% | 91.3% | 91.8% | 92.6% | 96.0% | 92.6% | 93.3% | 94.5% | 94.1% |
| 16 | 4 | (0.05, 0.10) | 10 | 89.4% | 89.5% | 90.5% | 96.0% | 90.6% | 91.9% | 93.7% | 92.5% | 89.4% | 90.7% | 95.9% | 90.8% | 91.9% | 93.7% | 94.8% |
| 16 | 4 | (0.05, 0.10) | 100 | 63.5% | 90.7% | 91.9% | 96.3% | 91.9% | 92.8% | 94.3% | 91.8% | 91.2% | 91.8% | 96.3% | 91.8% | 93.0% | 94.4% | 94.0% |
| 16 | 4 | (0.05, 0.15) | 10 | 85.8% | 89.2% | 90.3% | 96.1% | 90.4% | 91.6% | 93.8% | 91.2% | 89.5% | 90.9% | 95.8% | 91.0% | 92.0% | 93.8% | 93.9% |
| 16 | 4 | (0.05, 0.15) | 100 | 54.1% | 91.0% | 91.8% | 96.0% | 91.8% | 92.7% | 94.4% | 90.3% | 91.3% | 92.1% | 96.1% | 92.1% | 92.7% | 93.7% | 93.8% |
| 16 | 8 | (0.01, 0.05) | 10 | 89.9% | 90.1% | 91.1% | 98.6% | 91.2% | 92.0% | 93.1% | 91.0% | 90.0% | 91.0% | 98.5% | 91.2% | 91.9% | 92.9% | 94.8% |
| 16 | 8 | (0.01, 0.05) | 100 | 61.2% | 90.0% | 90.7% | 99.1% | 90.7% | 91.4% | 93.2% | 79.6% | 89.9% | 91.1% | 99.0% | 91.1% | 91.9% | 93.5% | 93.5% |
| 16 | 8 | (0.05, 0.10) | 10 | 85.8% | 90.3% | 91.1% | 98.7% | 91.2% | 91.9% | 93.2% | 90.0% | 90.3% | 91.0% | 98.8% | 91.0% | 91.9% | 93.2% | 94.4% |
| 16 | 8 | (0.05, 0.10) | 100 | 46.9% | 90.5% | 91.6% | 99.2% | 91.6% | 92.4% | 94.1% | 82.4% | 90.3% | 91.5% | 99.0% | 91.5% | 92.3% | 94.0% | 93.2% |
| 16 | 8 | (0.05, 0.15) | 10 | 79.6% | 90.2% | 90.8% | 98.6% | 90.8% | 91.4% | 93.4% | 87.0% | 90.3% | 90.8% | 98.5% | 90.9% | 91.9% | 93.4% | 94.2% |
| 16 | 8 | (0.05, 0.15) | 100 | 39.4% | 89.8% | 91.3% | 99.1% | 91.3% | 92.2% | 93.7% | 76.9% | 90.3% | 91.4% | 99.0% | 91.4% | 92.2% | 93.8% | 93.3% |
| 32 | 4 | (0.01, 0.05) | 10 | 92.7% | 93.5% | 93.7% | 95.5% | 93.7% | 93.9% | 94.7% | 93.7% | 93.5% | 93.9% | 95.5% | 93.9% | 94.2% | 94.9% | 94.9% |
| 32 | 4 | (0.01, 0.05) | 100 | 74.0% | 92.0% | 92.3% | 95.2% | 92.3% | 93.0% | 94.3% | 89.6% | 92.5% | 92.9% | 94.9% | 92.9% | 93.5% | 94.3% | 94.1% |
| 32 | 4 | (0.05, 0.10) | 10 | 91.0% | 94.1% | 94.3% | 96.1% | 94.4% | 94.7% | 95.3% | 94.1% | 93.8% | 94.1% | 96.2% | 94.1% | 94.4% | 95.5% | 95.2% |
| 32 | 4 | (0.05, 0.10) | 100 | 62.0% | 93.3% | 93.5% | 95.1% | 93.5% | 93.9% | 94.5% | 92.1% | 93.1% | 93.5% | 95.5% | 93.5% | 93.7% | 94.4% | 93.9% |
| 32 | 4 | (0.05, 0.15) | 10 | 87.8% | 93.6% | 94.3% | 96.4% | 94.3% | 94.4% | 95.1% | 93.1% | 93.7% | 94.0% | 96.1% | 94.0% | 94.5% | 95.6% | 95.4% |
| 32 | 4 | (0.05, 0.15) | 100 | 53.3% | 92.6% | 93.0% | 94.8% | 93.0% | 93.5% | 94.4% | 89.3% | 92.8% | 92.9% | 94.6% | 92.9% | 93.0% | 93.9% | 93.9% |
| 32 | 8 | (0.01, 0.05) | 10 | 90.1% | 93.5% | 93.9% | 96.9% | 93.9% | 94.5% | 95.1% | 91.5% | 93.6% | 94.0% | 97.0% | 94.1% | 94.5% | 95.0% | 94.3% |
| 32 | 8 | (0.01, 0.05) | 100 | 59.6% | 91.8% | 92.3% | 96.8% | 92.3% | 92.8% | 93.7% | 78.2% | 91.6% | 92.3% | 96.7% | 92.3% | 93.0% | 93.8% | 94.0% |
| 32 | 8 | (0.05, 0.10) | 10 | 86.2% | 94.0% | 94.2% | 96.9% | 94.2% | 94.6% | 95.0% | 90.2% | 94.0% | 94.2% | 97.0% | 94.2% | 94.6% | 95.2% | 94.7% |
| 32 | 8 | (0.05, 0.10) | 100 | 47.6% | 92.5% | 93.1% | 96.8% | 93.1% | 93.5% | 94.5% | 81.8% | 92.9% | 93.5% | 96.5% | 93.5% | 93.7% | 94.3% | 94.3% |
| 32 | 8 | (0.05, 0.15) | 10 | 82.3% | 93.7% | 94.0% | 96.5% | 94.0% | 94.5% | 94.8% | 88.4% | 93.9% | 94.2% | 96.5% | 94.2% | 94.4% | 94.8% | 94.3% |
| 32 | 8 | (0.05, 0.15) | 100 | 38.7% | 92.4% | 92.8% | 96.6% | 92.8% | 93.2% | 93.8% | 76.2% | 92.2% | 92.7% | 96.5% | 92.7% | 93.0% | 94.0% | 93.6% |

*Satterthwaite is not available for true models, the normal approxiamtion was used instead

**Table S5 Power comparison of selected scenarios under true LMM model and LMM model with CR3**

| I | S | WPICC (Control, Intervention) | K | Effect size | EXCH CR3 | NE CR3 | TRUE |
|---|---|---|---|---|---|---|---|
| | | **True model = NE + Random intervention** | | | | | |
| 8 | 4 | (0.01, 0.05) | 10 | 0.6 | 73.9% | 73.8% | 82.0% |
| 8 | 4 | (0.05, 0.10) | 10 | 0.6 | 62.9% | 63.6% | 72.0% |
| 8 | 4 | (0.05, 0.15) | 10 | 0.6 | 54.1% | 55.1% | 58.5% |
| 32 | 4 | (0.01, 0.05) | 10 | 0.3 | 89.4% | 89.5% | 91.1% |
| 32 | 4 | (0.05, 0.10) | 10 | 0.3 | 82.1% | 82.4% | 83.5% |
| 32 | 4 | (0.05, 0.15) | 10 | 0.3 | 74.0% | 74.6% | 75.1% |
| | | **True model = DTD + Random intervention** | | | | | |
| 8 | 4 | (0.01, 0.05) | 10 | 0.6 | 73.1% | 73.2% | 79.1% |
| 8 | 4 | (0.05, 0.10) | 10 | 0.6 | 62.7% | 62.9% | 67.3% |
| 8 | 4 | (0.05, 0.15) | 10 | 0.6 | 53.8% | 53.7% | 58.5% |
| 32 | 4 | (0.01, 0.05) | 10 | 0.3 | 90.4% | 90.7% | 91.0% |
| 32 | 4 | (0.05, 0.10) | 10 | 0.3 | 81.6% | 81.5% | 83.3% |
| 32 | 4 | (0.05, 0.15) | 10 | 0.3 | 73.1% | 72.2% | 75.1% |

| | | | | EXCH | | | NE | | | GEE Independent | | | GEE Exchangeable | | |
|---|---|---|---|---|---|---|---|---|---|---|---|---|---|---|---|
| | | | | | Coverage | | | Coverage | | | Coverage | | | Coverage | |
| I | S | WPICC (Control, Intervention) | K | Bias | CR2 | CR3 | Bias | CR2 | CR3 | Bias | KC | MD | Bias | KC | MD |
| | | | | **True model = NE + Random intervention** | | | | | | | | | | | |
| 8 | 4 | (0.01, 0.05) | 10 | -0.002 | 95.0% | 96.8% | -0.003 | 94.9% | 96.7% | -0.004 | 95.4% | 97.1% | -0.001 | 95.4% | 96.5% |
| 8 | 4 | (0.05, 0.10) | 10 | -0.001 | 94.6% | 97.0% | -0.001 | 94.5% | 96.8% | -0.004 | 94.8% | 97.2% | 0.000 | 97.0% | 96.6% |
| 8 | 4 | (0.05, 0.15) | 10 | -0.001 | 94.6% | 96.8% | 0.000 | 94.4% | 96.4% | 0.000 | 94.7% | 97.2% | -0.001 | 97.0% | 96.6% |
| 32 | 4 | (0.01, 0.05) | 10 | -0.001 | 94.1% | 95.0% | -0.001 | 93.9% | 94.6% | 0.001 | 93.7% | 94.3% | -0.002 | 95.1% | 94.9% |
| 32 | 4 | (0.05, 0.10) | 10 | -0.002 | 94.9% | 95.4% | -0.002 | 95.2% | 95.4% | -0.003 | 93.8% | 94.7% | 0.000 | 97.1% | 95.3% |
| 32 | 4 | (0.05, 0.15) | 10 | -0.002 | 94.8% | 95.3% | -0.002 | 94.2% | 95.2% | 0.001 | 93.2% | 94.3% | -0.002 | 97.2% | 95.3% |
| | | | | **True model = DTD + Random intervention** | | | | | | | | | | | |
| 8 | 4 | (0.01, 0.05) | 10 | -0.003 | 93.6% | 96.1% | -0.002 | 93.4% | 95.9% | 0.001 | 94.2% | 96.2% | -0.002 | 94.1% | 95.3% |
| 8 | 4 | (0.05, 0.10) | 10 | -0.002 | 93.9% | 95.9% | -0.003 | 93.8% | 96.0% | 0.002 | 94.8% | 96.8% | -0.002 | 96.0% | 95.6% |
| 8 | 4 | (0.05, 0.15) | 10 | -0.002 | 93.8% | 96.0% | -0.002 | 93.8% | 95.7% | 0.002 | 94.1% | 96.5% | 0.002 | 95.9% | 95.6% |
| 32 | 4 | (0.01, 0.05) | 10 | 0.002 | 94.8% | 95.5% | 0.002 | 94.8% | 95.5% | 0.000 | 94.9% | 95.3% | 0.001 | 95.6% | 95.6% |
| 32 | 4 | (0.05, 0.10) | 10 | 0.001 | 95.5% | 96.1% | 0.001 | 95.6% | 96.2% | 0.003 | 94.7% | 95.5% | -0.002 | 97.2% | 96.1% |
| 32 | 4 | (0.05, 0.15) | 10 | 0.000 | 95.4% | 96.4% | 0.001 | 95.3% | 95.6% | 0.001 | 95.1% | 95.8% | 0.000 | 97.3% | 96.4% |

Table S6 Comparison of coverage probabilities under linear mixed model and generalized estimating equation

**KC: Kauermann and Carroll**

**MD: Mancl and DeRouen**

**Table S7 Type I error comparison under linear mixed model and permutation test**

| I | S | WPICC (Control, Intervention) | K | EXCH | | NE | | Permutation |
|---|---|---|---|---|---|---|---|---|
| | | | | CR2 | CR3 | CR2 | CR3 | |
| | | **True model = NE + Random intervention** | | | | | | |
| 8 | 4 | (0.01, 0.05) | 10 | 5.1% | 3.2% | 5.2% | 3.3% | 6.4% |
| 8 | 4 | (0.05, 0.10) | 10 | 5.5% | 3.0% | 5.6% | 3.3% | 7.6% |
| 8 | 4 | (0.05, 0.15) | 10 | 5.5% | 3.2% | 5.6% | 3.6% | 8.8% |
| 32 | 4 | (0.01, 0.05) | 10 | 6.0% | 5.0% | 6.2% | 5.4% | 8.4% |
| 32 | 4 | (0.05, 0.10) | 10 | 5.1% | 4.6% | 4.8% | 4.6% | 8.6% |
| 32 | 4 | (0.05, 0.15) | 10 | 5.2% | 4.8% | 5.8% | 4.9% | 9.6% |
| | | **True model = DTD + Random intervention** | | | | | | |
| 8 | 4 | (0.01, 0.05) | 10 | 6.4% | 3.9% | 6.6% | 4.1% | 6.6% |
| 8 | 4 | (0.05, 0.10) | 10 | 6.1% | 4.1% | 6.3% | 4.0% | 7.8% |
| 8 | 4 | (0.05, 0.15) | 10 | 6.3% | 4.0% | 6.3% | 4.3% | 9.2% |
| 32 | 4 | (0.01, 0.05) | 10 | 5.2% | 4.5% | 5.2% | 4.6% | 7.0% |
| 32 | 4 | (0.05, 0.10) | 10 | 4.5% | 3.9% | 4.5% | 3.9% | 6.6% |
| 32 | 4 | (0.05, 0.15) | 10 | 4.6% | 3.6% | 4.4% | 3.9% | 7.6% |

**Figure S1 The percentage error of model-based standard error to empirical standard error under both the true model and misspecified models with CR0 and CR3 when the data was generated from DTD-RI model**

CR0 is the standard robust variance estimator, and CR3 closely approximates the leave-one-cluster-out jackknife resampling estimator.

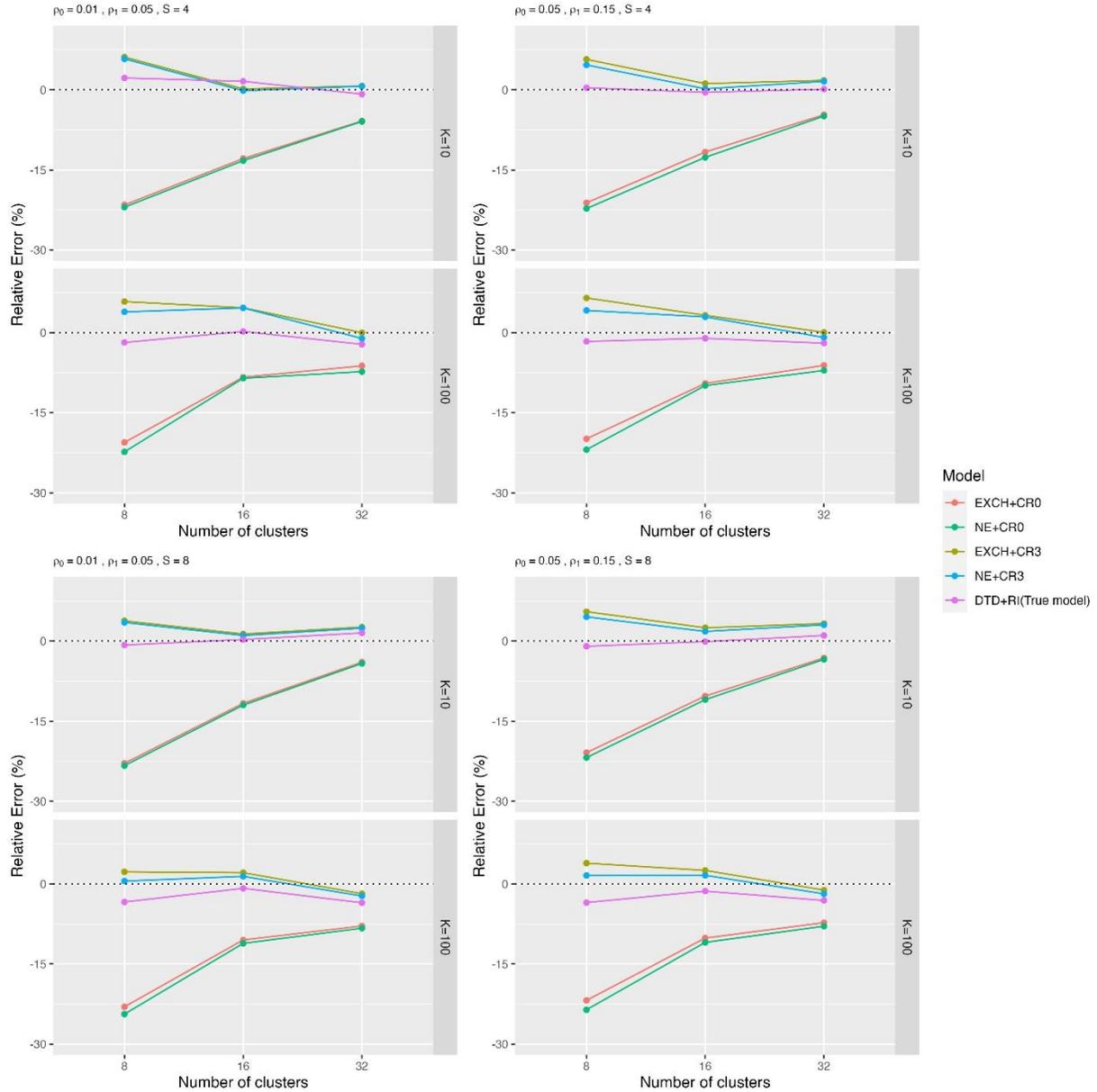

**Figure S2 The percentage error of model-based standard error to empirical standard error under both the true model and misspecified models with CR0 and CR3 when the data was generated from NE-RI model**

CR0 is the standard robust variance estimator, and CR3 closely approximates the leave-one-cluster-out jackknife resampling estimator.

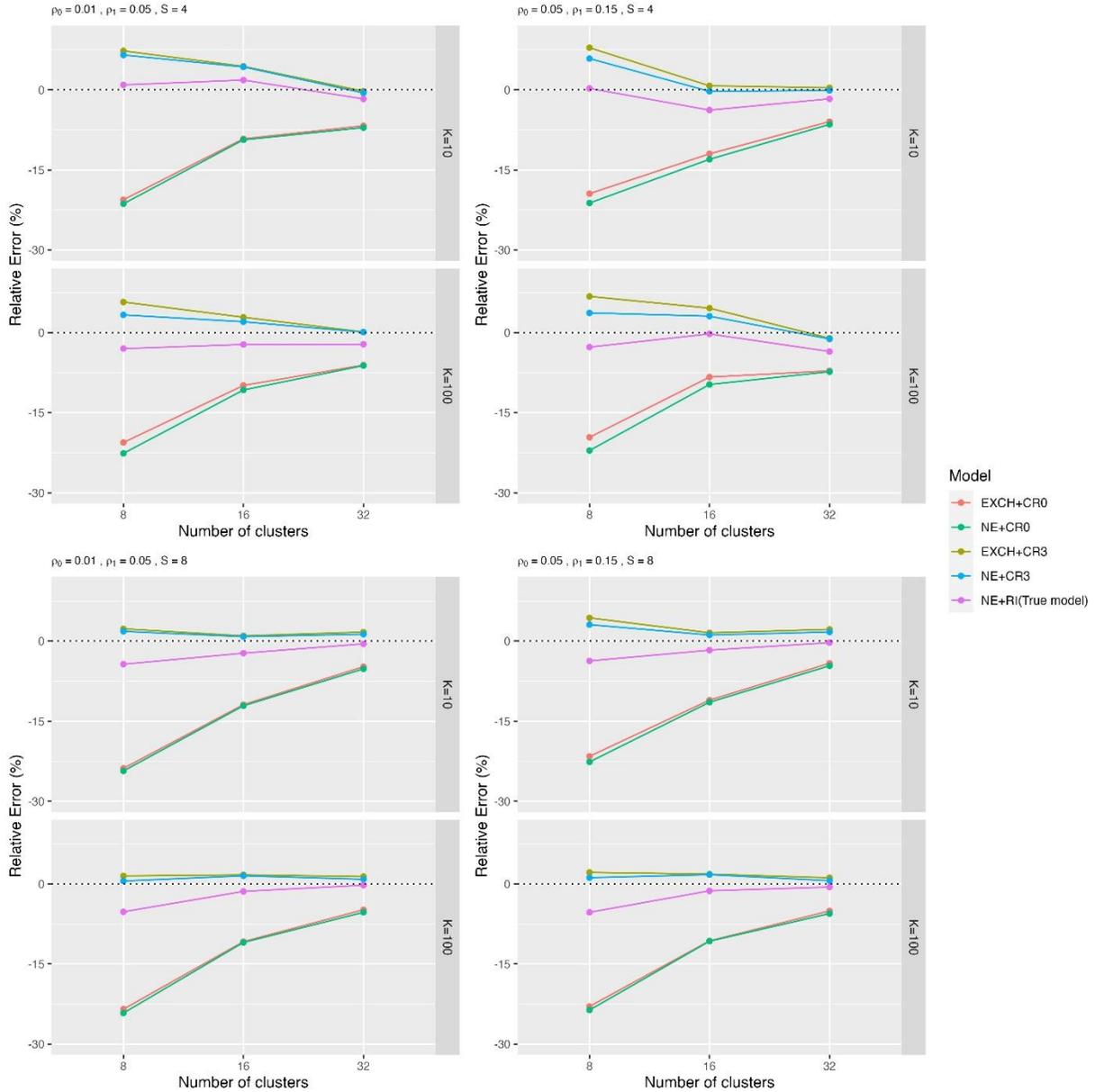